\documentclass[aps,floatfix,onecolumn,a4paper,noshowpacs, nofootinbib,superscriptaddress,11pt]{revtex4}
\usepackage{float,xcolor,upgreek,yfonts,tikz}
\usepackage{color}
\usepackage{xcolor}
\usepackage{bm}
\usepackage{comment}
\usepackage{graphicx,bigints}
\usepackage{graphicx,float}\usepackage[all]{xy}
\usepackage{amsmath,amssymb,upgreek}

   \usepackage{caption}

\usepackage{subcaption}
\usepackage{physics}
\usepackage{tensor}
\usepackage{amssymb}

\usepackage{alphabeta}
 \usepackage{cancel}
\usepackage{dcolumn}
\usepackage{textgreek}
\usepackage{adjustbox}
\usepackage{multirow} 
\usepackage{amsmath}
\allowdisplaybreaks
\usepackage{units}
\usepackage{overpic}

\newcommand{\beq}{\begin{eqnarray}}
\newcommand{\eeq}{\end{eqnarray}}
\newcommand{\be}{\begin{equation}}
\newcommand{\ee}{\end{equation}}

\newcommand{\bea}{\begin{eqnarray}}
\newcommand{\eea}{\end{eqnarray}}

\newcommand{\ba}{\begin{eqnarray}}
\newcommand{\ea}{\end{eqnarray}}

\usepackage[colorlinks,hyperindex,unicode]{hyperref}
\definecolor{green1}{RGB}{0,128,0} 
\hypersetup{hidelinks,backref=true,pagebackref=true,hyperindex=true,colorlinks=true,breaklinks=true,urlcolor= blue}
\hypersetup{%
  colorlinks = true,
  linkcolor  = blue,
  citecolor = cyan,
}
\usepackage{bookmark,textgreek}
\usepackage{hyperref,color,xcolor}
\hypersetup{hidelinks,hyperindex=true,colorlinks=true,breaklinks=true,urlcolor= blue}
\hypersetup{%
  colorlinks = true,
  linkcolor  = blue
}
\newcommand\orcidkelvin{{\href{https://orcid.org/0000-0002-3822-9818}{\orcidicon}}}

\newcommand\orcidronaldo{{\href{https://orcid.org/0000-0001-7004-4656}{\orcidicon}}}

\newcommand\orcidrogerio{{\href{https://orcid.org/0000-0001-7848-5472}{\orcidicon}}}

\newcommand\orcidroldao{{\href{https://orcid.org/0000-0003-3978-532X}{\orcidicon}}}
\newcommand{\orcidicon}{%
	\begin{tikzpicture}
	\draw[lime, fill=lime] (0,0)
		circle [radius=0.16]
		node[white] {{\fontfamily{qag}\selectfont \tiny ID}};
	\draw[white, fill=white] (-0.0625,0.095)
		circle [radius=0.007];
	\end{tikzpicture}	\hspace{-2mm}
}

\usepackage{graphicx}
\usepackage{dcolumn}
\usepackage{bm}
\usepackage{multirow}
\usepackage{caption}
\usepackage{subcaption}
\usepackage{comment} 
\usepackage{float}

\begin{document}

\title{Gravitational decoupling and regular hairy black holes: Geodesic stability, quasinormal modes, and thermodynamic properties}

\author{R. C. de Paiva\orcidronaldo}
\affiliation{Physics Department, São Paulo State University, 12516-410, Guaratingueta, Brazil}
\email{ronaldo.paiva@unesp.br}

\author{K. S. Alves\orcidkelvin}
\affiliation{Physics Department, São Paulo State University, 12516-410, Guaratingueta, Brazil}
\email{kelvin.santos@unesp.br}

\author{R. T. Cavalcanti\orcidrogerio}
\affiliation{Physics Department, São Paulo State University, 12516-410, Guaratingueta, Brazil}
\affiliation{Institute of Mathematics and Statistics, Rio de Janeiro State University, 20550-900, Rio de Janeiro, Brazil}
\email{rogerio.cavalcanti@ime.uerj.br}

\author{R. da Rocha\orcidroldao}
\affiliation{Center of Mathematics, Federal University of ABC, 09210-580, Santo Andr\'e, Brazil}
\email{roldao.rocha@ufabc.edu.br}

\medbreak
\begin{abstract} 
The stability of geodesic orbits around a regular hairy black hole, in the gravitational decoupling setup, is investigated by employing Lyapunov exponents, which quantify the divergence rate of nearby trajectories in dynamical systems. Both timelike and null geodesics are addressed, probing the effect of the hair parameter on orbital stability. Deviations from the Schwarzschild solution have a significant influence on orbit stability, potentially providing observational signatures. We compute the quasinormal modes of regular hairy black holes and discuss their thermodynamic properties, contrasting the Rényi and Bekenstein-Hawking entropy prescriptions. The results can provide insight into gravitational dynamics in the strong-field regime and potentially contribute to ongoing developments in modified gravity theories.
\end{abstract}



\maketitle

\section{Introduction}
\label{sec0}
The groundbreaking detection of gravitational waves (GWs), particularly from the final moments of merging binary systems, has profoundly enhanced our understanding of gravity, especially in its strongest regime. This significant achievement has not only marked a turning point in astrophysics but has also provided experimental validation of solutions to Einstein's field equations and their extensions \cite{LIGOScientific:2016lio,LIGOScientific:2019fpa}.
One of the pivotal methods in this field is the gravitational decoupling (GD) technique, which is an effective tool for deriving self-gravitating compact stellar configurations from the well-established solutions of general relativity (GR). In the GD setup the usual matter source $T_{\mu\nu}$ is coupled to an additional source $\Uptheta_{\mu\nu}$ \cite{Ovalle:2017fgl,Ovalle:2018gic}, representing a new gravitational sector encoding hairy fields, possibly including tidal and gauge charges, also accommodating any possibility involving effects of dark matter and dark energy, moduli fields and hidden sectors beyond the Standard Model, and quantum corrections to gravity. Ref. \cite{Maurya:2024zao} employed the GD to investigate the Bose--Einstein condensation in the dark matter context. The possibility of using the additional source in GD to report dark matter and dark energy was addressed in Refs. \cite{LinaresCedeno:2019aul,Tello-Ortiz:2020ydf}. Also, dark matter halos and dark matter signatures from GD compact stars were explored in Refs. \cite{Maurya:2025jzj,Pradhan:2024hne,Yousaf:2024src,Maurya:2024ylr}.    
Refs. \cite{daRocha:2020gee,DaRocha:2019fjr} utilized the GD in the context of the holographic entanglement entropy of hairy black holes, whereas axion stars were scrutinized through the GD, emulating mini-massive compact halo objects  \cite{Casadio:2023mgl}.

Relevant compact stellar configurations derived from the GD technique have been reported ~\cite{Estrada:2019aeh, Gabbanelli:2019txr, Leon:2023nbj,Ramos:2021drk, Rincon:2019jal, Panotopoulos:2018law, Singh:2019ktp, Ovalle:2018ans,PerezGraterol:2018eut, Heras:2018cpz, Torres:2019mee}. Notably, relativistic approaches for nuclear interactions suggest that, under conditions of extreme density, the interiors of stars may exhibit significant anisotropies. 
The GD explores the anisotropic distributions of matter within stars, which facilitates the formulation of analytical solutions to Einstein's field equations. This is achieved by incorporating diverse configurations for the energy-momentum tensor, allowing for a richer understanding of how gravity operates in extreme conditions \cite{Tello-Ortiz:2021kxg, Zubair:2023cvu, Bamba:2023wok, Maurya:2023uiy, Iqbal:2025xqf,Contreras:2021xkf, Sharif:2020lbt}. Additionally, the GD method can be examined from the point of view of gauge/gravity duality,  
testing the physical viability of quantum corrections to black holes, particularly in the infrared limit  \cite{Meert:2020sqv,daRocha:2014dla}.  
Hairy black holes incorporate additional fields beyond the original no-hair theorem \cite{Babichev:2013cya, Sotiriou:2013qea, Antoniou:2017acq}, playing a prominent role in the implications of alternative theories of gravity and quantum gravity phenomenology \cite{Hayward:2005gi}.
Ref. \cite{Ovalle:2020kpd} introduced hairy GD black holes, which have been comprehensively applied to relevant developments with observational signatures \cite{Liang:2024xif,Avalos:2023ywb,Zhang:2022niv,Ditta:2023arv,Mahapatra:2022xea,Albalahi:2024vpy, Misyura:2024fho}. 
 In this context, quasinormal modes and GW analysis,  emitted from GD hairy  solutions, include relevant observational aspects  \cite{Cavalcanti:2016mbe,Cavalcanti:2022cga,Yang:2022ifo,Li:2022hkq,Cavalcanti:2022adb,Guimaraes:2025jsh,Avalos:2023jeh,Tello-Ortiz:2024mqg, Meng:2025glf}. Ref. \cite{daRocha:2017cxu}
used the GD to scrutinize glueball dark matter models and their collapse that originate GD compact stellar configurations, while the GD can also describe sub-Planckian black holes and black hole remnants  \cite{Casadio:2017sze}.
Regular black holes have gained attention as possible resolutions of physical singularities, also encompassed in extensions of GR  \cite{Poisson:1990eh,Bonanno:2022jjp,Carballo-Rubio:2022kad}. 
Refs.  \cite{Tang:2024txx,Stashko:2024wuq,Konoplya:2024hfg, Balart:2023odm} discuss quasinormal modes associated with regular black hole solutions, which avoid singularities through modifications encoded by black hole hair. 
Ref. \cite{Ovalle:2023ref} introduced regular hairy black holes, using the tensor vacuum in GD. 

As the gravitational field and related phenomena are extremely intense in the vicinity of black holes, the strong-field regime is a significant test of GR and its extensions. A usual phenomenon consists of the accretion of matter from an accretion disk. Massive and massless particles orbit black holes, as evidenced by observational data of the M87 black hole \cite{EventHorizonTelescope:2022wkp}. One of our primary objectives is to characterize the orbits of massive and massless particles around a GD regular hairy black hole. We apply the Lyapunov exponents in the context of dynamical systems to quantify the instability of geodesic orbits. Lyapunov exponents quantify how quickly nearby trajectories diverge, providing a robust framework for analyzing orbital stability \cite{Cardoso:2008bp,Giri:2022zhf,Mondal:2020uwp}. Ref. \cite{Olmo:2023lil} used Lyapunov exponents to investigate shadows and photon rings of regular black holes in modified gravity. In this work, we investigate geodesic stability around a regular hairy black hole using Lyapunov exponents. Both timelike and null geodesics are examined, together with their stability conditions and the effects of the hair parameter on orbital dynamics. By doing so, we aim to elucidate how deviations from Schwarzschild black holes, implemented by the GD regular hairy black holes,  affect the stability and observable properties of orbits, thereby contributing to a deeper understanding of gravitational phenomena in modified gravity theories.
 
 This work is organized as follows: Sec. \ref{sec1} is devoted to briefly introducing the GD method and, subsequently, a family of regular hairy black holes, besides fixing the notation to be used throughout the paper.  In Sec. \ref{sec2}, we analyze the geodesic stability around regular hairy black holes, computing the corresponding Lyapunov exponents for null and time-like geodesics.  We conclude the section by examining whether gauging the hairy parameter reproduces the results of Kerr rotating black holes in Sec. \ref{sec4}. In Sec. \ref{sec3}, we study the quasinormal modes of regular hairy black holes for scalar, electromagnetic, and gravitational perturbations, as well as the quasinormal modes in the eikonal limit. Thereafter, the regular hairy black hole thermodynamics is scrutinized in Sec. \ref{sec5}, the Hawking temperature is calculated, and both the  Rényi and the Bekenstein-Hawking entropies are reported. Sec. \ref{sec6} is dedicated to calculating the emission rate of regular hairy black holes, and our conclusions and final remarks are presented in  Section \ref{sec7}.
 
\section{Regular hairy black hole through gravitational decoupling}

\label{sec1}
In this section, a regular hairy black hole metric, obtained via the GD method, is derived. Hereon, natural units will be used. One starts from Einstein's field equations
\begin{equation}
    G_{\mu\nu} \;=\; 8\pi\,\mathring{T}_{\mu\nu},
\end{equation}
where $G_{\mu\nu} = R_{\mu\nu} - \frac{1}{2}R\,g_{\mu\nu}$ stands for  the Einstein tensor and $\mathring{T}_{\mu\nu}$ denotes the effective energy-momentum tensor. It can be split off as 
\begin{equation}\label{tmn00}
    \mathring{T}_{\mu\nu} \;=\; T_{\mu\nu} + \,\Uptheta_{\mu\nu}.
\end{equation}
Here, $T_{\mu\nu}$ is the energy-momentum tensor associated with standard matter, typically describing a perfect fluid, whereas $\Uptheta_{\mu\nu}$ accounts for an additional sector, sourcing gravity. {According to the geometric deformation prescription \cite{Ovalle:2016pwp} and its current implementation within the GD framework, the additional source $\Theta_{\mu\nu}$ plays the role of an effective gravitational sector responsible for deforming the seed geometry. In the GD approach, the splitting (\ref{tmn00}), represents the sum between the seed source and the extra anisotropic sector that gravitationally decouples from the seed configuration. The essential point is that the GD formalism is constructed at the level of Einstein's field equations, keeping the Einstein tensor and the Einstein--Hilbert gravitational sector unchanged, while the effects of new physics are effectively absorbed into the additional source term $\Theta_{\mu\nu}$. Therefore, the nature of $\Theta_{\mu\nu}$ is intentionally left unspecified to preserve the generality of the method. It may represent either an additional matter sector minimally coupled to gravity or effective anisotropic corrections, and it can also incorporate low-energy modifications arising from extensions of GR, including string-inspired effects, higher-order curvature and semiclassical quantum-gravitational corrections, provided these contributions can be consistently recast as effective sources in the Einstein field equations, where any geometric modification is absorbed into an effective energy-momentum tensor. Consequently, $\Theta_{\mu\nu}$ parametrizes deviations from the seed solution while preserving the standard GR structure of the gravitational sector. Therefore, the additional source effectively parametrizes deviations from the seed solution without modifying the Einstein--Hilbert structure of the gravitational sector. No explicit non-minimal couplings between curvature invariants and matter fields are introduced in the present construction. Hence, the thermodynamic analysis to be implemented in Sec. \ref{sec5} is consistent with the standard Einstein gravity framework, and the Bekenstein--Hawking entropy formula holds. }

From the Bianchi identities, the total source term  necessarily satisfies the conservation law  \cite{Ovalle:2017fgl,Ovalle:2020kpd}:
\begin{equation}\label{bianc}
    \nabla^\mu\mathring{T}_{\mu\nu} \;=\; \nabla^\mu\left(T_{\mu\nu}+\,\Uptheta_{\mu\nu}\right) \;=\; 0.
\end{equation}
The pivotal idea underlying the GD procedure is to start with a known solution of Einstein's field equations and introduce a deformation of the metric, such that the field equations can be separated into two systems. The first one corresponds to the original matter sector $T_{\mu\nu}$, which is called seed source in the GD setup, whereas the decoupled one governs the deformation process and is implemented by the $\Uptheta_{\mu\nu}$ term. Assuming a general spherically symmetric metric,
\begin{equation}\label{deformed_metric}
   {d}s^2 = -e^{\xi(r)}{d}t^2 + e^{\zeta(r)}{d}r^2 + r^2 {d}\Omega^2,
\end{equation}
Einstein's field equations, therefore, yield
\begin{align}
    8\pi\left( T\indices{^{0}_{\;0}} + \Uptheta\indices{^{0}_{\;0}}\right) &= -\frac{1}{r^2} -\left(\zeta^{\prime}  - \frac{1}{r}\right)\frac{e^{-\zeta}}{r},\label{00}\\
    8\pi\left( T\indices{^{1}_{\;1}} + \Uptheta\indices{^{1}_{\;1}}\right) &=  -\frac{1}{r^2} +\left(\xi^{\prime}  + \frac{1}{r}\right)\frac{e^{-\zeta}}{r}, \label{11}\\
    8\pi\left( T\indices{^{2}_{\;2}} + \Uptheta\indices{^{2}_{\;2}}\right) &={\left[2r \xi^{\prime\prime} + r\left(\xi^{\prime}\right)^{2} - {\left(r\zeta^{\prime} - 2\right)}\xi^{\prime}  - 2\zeta^{\prime}\right]} \frac{e^{-\zeta}}{4r}.\label{22}
\end{align}
One can consider a spherically symmetric solution to the seed source $T_{\mu\nu}$, as 
\begin{equation}\label{ndeformed_metric}
   {d}s^2 = -e^{\chi(r)}{d}t^2 + e^{\uplambda(r)}{d}r^2 + r^2 {d}\Omega^2.
\end{equation}
Now, one can induce a deformation on the functions $\chi(r)$ and $\uplambda(r)$, in such a way that
\begin{subequations}
\begin{eqnarray}
\label{gd1}
\chi(r)
&\mapsto &
\xi(r) = \chi(r) +\alpha f_1(r),
\\
\label{gd2}
e^{-\uplambda(r)}
&\mapsto &
e^{-\zeta(r)} = e^{-\uplambda(r)} +\alpha f_2(r),
\end{eqnarray}
\end{subequations}
where $f_1$ and $f_2$ are respectively the geometric deformations for the temporal and radial metric components, parametrized by the coupling constant $\alpha$. Substituting them in Einstein's field equations, the resulting system decouples as
\begin{align}
    8\pi T\indices{^{0}_{\;0}}  &= -\frac{1}{r^2} -\left(\uplambda^{\prime}  - \frac{1}{r}\right)\frac{e^{-\uplambda}}{r},\label{00}\\
        8\pi T\indices{^{1}_{\;1}}  &=  -\frac{1}{r^2} +\left(\chi^{\prime}  + \frac{1}{r}\right)\frac{e^{-\uplambda}}{r}, \label{11}\\
        8\pi T\indices{^{2}_{\;2}}  &={\left[2r \chi^{\prime\prime} + r\left(\chi^{\prime}\right)^{2} - {\left(r\uplambda^{\prime} - 2\right)}\chi^{\prime}  - 2\uplambda^{\prime}\right]} \frac{e^{-\uplambda}}{4r},\label{22}
\end{align}
for the standard Einstein's field equations with seed source term $T_{\mu\nu}$, and a second set of equations for the additional source $\Uptheta_{\mu\nu}$ given by
\begin{align}
8\pi \Uptheta\indices{^{0}_{\;0}} &= - \alpha \frac{f_2}{r^2} - \alpha \frac{f_2'}{r},  \label{m00}\\
8\pi \Uptheta\indices{^{1}_{\;1}} - \alpha \mathcal{Z}_1 &= \alpha f_2 \left( \frac{1}{r^2} + \frac{\xi'}{r} \right), \label{m11} \\
8\pi \Uptheta\indices{^{2}_{\;2}} - \alpha \mathcal{Z}_2 &= \frac{\alpha f_2}{4} \left( 2 \xi'' + \xi'^2 + \frac{2\xi'}{r} \right)
+ \frac{\alpha f_2'}{4} \left( \xi' + \frac{2}{r} \right), \label{m22}
\end{align}
where
\begin{align}
\mathcal{Z}_1 &= e^{-\uplambda} \frac{f_1'}{r}, \\
 \mathcal{Z}_2 &= \frac14e^{-\uplambda} \left( 2 f_1'' + \alpha f_1'^2 + \frac{2 f_1'}{r} + 2 f_1' \chi' - \uplambda' f_1' \right).
\end{align}
The perfect fluid model allows the identification of the effective energy density, the radial pressure, and the tangential pressure, respectively, as
\begin{align}\label{pian}
    \mathring{\upepsilon} &\equiv- T\indices{^{0}_{0}} - \Uptheta\indices{^{0}_{0}} = \upepsilon  + \mathcal{E},\\
    \mathring{p}_{r} &\equiv T\indices{^{1}_{1}} + \Uptheta\indices{^{1}_{1}} = p_r + \mathcal{P}_r,\\
    \mathring{p}_{t} &\equiv T\indices{^{2}_{2}} + \Uptheta\indices{^{2}_{2}} = p_\Uptheta + \mathcal{P}_\Uptheta.\label{pia2}
\end{align}
Anisotropy naturally sets in, since $ \mathring{p}_{t}- \mathring{p}_{r} \neq0$, in general. Therefore, the system (\ref{pian}) -- (\ref{pia2})  is suitable to describe an anisotropic fluid.
Consequently, one must deal with a system of three equations and five unknowns, namely $f_1, f_2, \mathcal{E}, \mathcal{P}_r$, and $\mathcal{P}_\Uptheta$. Additional conditions should thus be imposed.

The next step consists of selecting a seed solution for the system described by Eqs. \eqref{00} - \eqref{22}. Choosing the Schwarzschild metric, $e^{\chi(r)} = e^{-\uplambda(r)} = 1-\frac{2M_s}{r}$, where $M_s$ denotes the seed mass, and demanding that the deformed metric \eqref{ndeformed_metric}  has an associated  horizon surface $r=r_h$ which is simultaneously a Killing horizon\footnote{A Killing horizon is a null hypersurface defined as the locus where the Killing vector field is light-like.} for the metric (\ref{deformed_metric}) $[e^{\xi(r_h)} = 0]$ and a causal horizon $[e^{\zeta(r_h)} = 0]$, it is sufficient to impose the condition 
\beq\label{kash}
e^{\xi(r)} = e^{\zeta(r)}.\eeq Doing it, Eq.  \eqref{deformed_metric} takes the form
\begin{equation}\label{new_deformed_metric}
   {d}s^2 = -\left(1-\frac{2M_s}{r}\right)h(r)dt^2 + \left(1-\frac{2M_s}{r}\right)^{-1}\frac{{d}r^2}{h(r)} + r^2 {d}\Omega^2,
\end{equation}
 with $h(r) = e^{\alpha f_1(r)}$. 
Ref.  \cite{Ovalle:2023ref} considered the tensor-vacuum, for which the seed source vanishes,  $T_{\mu\nu}=0$,  yielding
\begin{equation}
{\cal P}_{r}
	=
	-{\cal E}.
	\label{seeso}
\end{equation}
As a consequence, for positive values of the energy density, only a negative radial pressure is allowed. 
Eqs. \eqref{bianc} and \eqref{seeso}
leads to 
\begin{eqnarray}
	\label{eq23}
	{\cal P}_r'=
	\frac{2}{r}\left({{\cal P}_\Uptheta}-{\cal P}_r\right),
\end{eqnarray}
 corresponding to hydrostatic equilibrium, preventing the additional gravitational source $\Uptheta_{\mu\nu}$ from collapsing into the central singularity associated with the seed Schwarzschild metric  \cite{Ovalle:2023ref}.

Upon performing the necessary algebraic manipulations, the metric is completely specified by enforcing the weak energy condition (WEC), leading to \cite{Ovalle:2023ref}
\begin{equation}
\label{ge}
8\pi\,{\cal E}
=
\frac{\alpha}{\ell^2}\,e^{-r/\ell}
\ ,
\end{equation}
where $\ell$ is a constant with dimensions of a length. 
The real hairy parameter $\alpha$ is introduced in Eq.~\eqref{ge} to recover the seed vacuum solution in the limit $\alpha\to 0$.
The  asymptotically flat regular hairy black hole metric is given by 
\begin{align}\label{hbr}
e^{\xi(r)} = e^{-\chi(r)} = 1-\frac{2M}{r} +  \frac{2e^{-\alpha\frac{r}{M}}}{Mr}  \left( M^2 + {M\alpha r}  +\frac{\alpha^2r^2}{2} \right),
\end{align}
where $\alpha\ell = M$ is the ADM mass. 
The scalar curvature, the Ricci squared, and the Kretschmann scalar for the metric \eqref{hbr} were shown to assume finite values \cite{Ovalle:2023ref} when $r\to 0$. Therefore, such a solution has no curvature singularities.

\begin{figure}[H]
\centering
\includegraphics[width=0.8\linewidth]{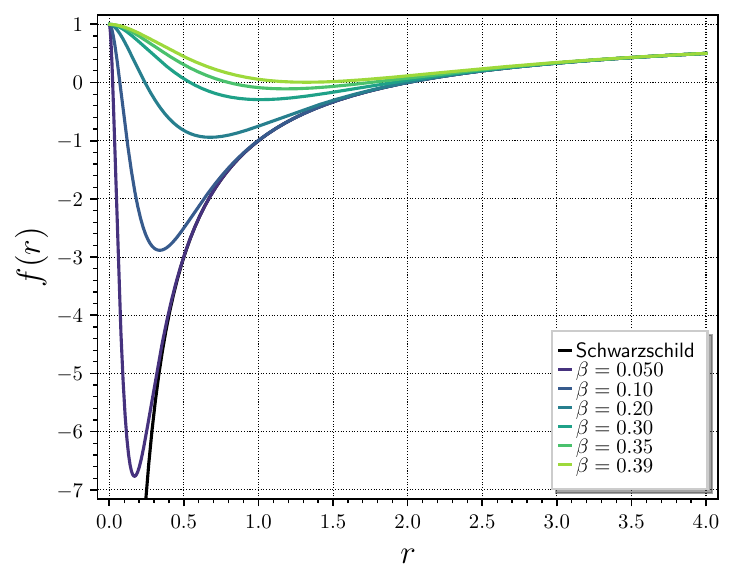}
\caption{\footnotesize \footnotesize Metric coefficient \eqref{hrbh} of the regular hairy black hole \eqref{main_metric}. The plot displays the existence of two horizons, corresponding to the zeroes of $f$, for $0<\beta<0.39$. The radial coordinate is rescaled and expressed in units of the Schwarzschild mass.}\label{hor11}
\end{figure}

 For convenience, we introduce the hairy parameter as $\beta = \alpha^{-1}$, leading to the following metric
\begin{equation}\label{main_metric}
 ds^2 = -f(r,\beta)dt^2 + f(r,\beta)^{-1}dr^2 + r^2d\Omega^2,
\end{equation}
with 
\begin{equation}\label{hrbh0}
f(r,\beta) \equiv e^{\xi(r,\beta)} = e^{-\chi(r,\beta)} = 1 -  \frac{2M}{r}  +  \frac{2e^{-\frac{r}{M\beta}}}{Mr}  \left( M^2 + \frac{Mr}{\beta}  +\frac{r^2}{2\beta^2} \right).
\end{equation}
Throughout this work, we shall express the radial coordinate in units of the Schwarzschild mass, redefining $r/M \to r$. Under this rescaling, the metric takes the following form
\begin{equation}\label{hrbh}
f(r,\beta)  = 1 -  \frac{2}{r}  +  \frac{2e^{-\frac{r}{\beta}}}{r}  \left( 1 + \frac{r}{\beta}  +\frac{r^2}{2\beta^2} \right).
\end{equation}
Notice that the following limits hold:
\begin{align}
\lim_{\beta \to 0} f(r,\beta) &= 1 - \frac{2}{r}, \qquad\qquad\quad
\lim_{\beta \to \infty} f(r,\beta) = 1.
\end{align}
Thus, depending on the value of the hairy parameter $\beta$, the solution \eqref{main_metric}, with coefficients \eqref{hrbh}, approaches either the Schwarzschild spacetime or the flat Minkowski spacetime. Regarding the event horizon radius $r_h$, defined by $f(r_h) = 0$, varying $\beta$ reveals three distinct possibilities, separated by a critical value $\beta_{\text{crit}} \approx 0.39$, for which the configuration becomes extremal and admits a single event horizon. If $\beta < \beta_{\text{crit}}$, there are two horizons. When $\beta = \beta_{\text{crit}}$, an extremal configuration encodes a single horizon, and for $\beta > \beta_{\text{crit}}$ there are no horizons.
The horizons that appear in this solution are the Cauchy horizon \cite{Poisson:1990eh}, which is a special type of surface in spacetime that marks the boundary beyond which the initial conditions no longer uniquely determine the future evolution of Einstein’s equations, and the event horizon.
The event horizon of the regular hairy black hole \eqref{main_metric}, with coefficients \eqref{hrbh}, is depicted as a function of the radial coordinates, for several values of the hair parameter $\beta$ in Fig. 
\ref{hor11}. 

\section{Geodesic stability around regular hairy black hole via Lyapunov exponents}
\label{sec2}

Lyapunov exponents are fundamental tools in the study of dynamical systems, widely used to characterize the sensitivity of trajectories to initial conditions. They quantify how rapidly nearby trajectories diverge (or converge) over time, providing a rigorous criterion for the stability or chaotic nature of dynamical evolution. In particular, the largest Lyapunov exponent serves as an indicator of the average exponential rate of divergence of infinitesimally close trajectories in phase space. Lyapunov exponents have been applied in cosmology and GR~\cite{Shukla:2024tkw,Das:2023ess,Du:2024uhd,Gogoi:2024akv,Barreto:2022ohl}.

Let us consider a trajectory $x(t)$ that is a solution of a $4$-dimensional autonomous dynamical system described by
\begin{equation}
    \frac{dx(t)}{dt} = y(x),
\end{equation}
and a nearby perturbed trajectory represented by a deviation vector $\varsigma(t)$, given by $\varsigma(t) = x(t) - x_0$, where $x_0$ is the initial condition associated with $x(t)$. Under the assumption that the perturbation remains small, the evolution of $\varsigma(t)$ can be approximated by the linearized variational equation
\begin{equation}
\label{linearizedeq}
    \frac{d\varsigma(t)}{dt} = \mathbb{L}(x)\, \varsigma(t),
\end{equation}
where $\mathbb{L}(x)$ is the Jacobian matrix of the vector field $y(x)$. The formal solution to Eq.~\eqref{linearizedeq} is given by $\varsigma(t) = A^t \varsigma(0)$, where $A^t$ is the evolution operator that maps the initial perturbation $\varsigma(0)$ to the perturbed state $\varsigma(t)$. The principal Lyapunov exponent $\lambda$ is defined as~\cite{sano1985measurement}
\begin{equation}
    \lambda = \lim_{t \to \infty} \frac{1}{t} \ln \left( \frac{\|\varsigma(t)\|}{\|\varsigma(0)\|} \right),
\end{equation}
where $\|\cdot\|$ denotes a norm on the tangent space of the phase space, typically induced by a suitable metric.

In the context of GR, Lyapunov exponents have been used to study the (in)stability of particle orbits in curved spacetimes, including those around black holes. Of particular interest is the stability of circular geodesics, which plays a critical role in understanding the dynamics of accretion disks and GW emission. For a non-spinning test particle moving in a static and spherically symmetric spacetime, the radial equation of motion can be expressed in terms of an effective potential. Near a circular orbit, the second derivative of the effective potential determines stability criteria. Positive curvature corresponds to stability, while negative curvature indicates instability.
In such spacetimes, one can derive expressions for the Lyapunov exponents associated with small radial perturbations around circular orbits. The proper-time Lyapunov exponent $\lambda_p$, which measures the rate of divergence in the particle's proper frame, is given by~\cite{Giri:2022zhf,Cardoso:2008bp,Mondal:2020uwp,Du:2024uhd,Das:2023ess}
\begin{equation}
    \lambda_p = \pm \sqrt{\frac{(\dot{r}^2)'}{2}}.
\end{equation}
Hereon, a prime (\;\;$'$\;\;) now denotes differentiation with respect to the radial coordinate $r$, and an overdot (\;\;$\dot{}$\;\;) denotes differentiation with respect to the affine parameter $\tau$, often taken to be the proper time for timelike geodesics. 
The coordinate-time Lyapunov exponent $\lambda_0$, measured with respect to the asymptotic time coordinate, reads
\begin{equation}
    \lambda_0 = \pm \sqrt{\frac{(\dot{r}^2)'}{2 \dot{t}^2}}.
\end{equation}
As usual, the $\pm$ sign will be omitted, and only positive Lyapunov exponents are considered. A circular orbit is stable when the Lyapunov exponent $\lambda$ is imaginary, unstable when $\lambda$ is real, and marginally stable (or a saddle point) when $\lambda$ equals zero.

For completeness, before deriving the Lyapunov exponents associated with timelike and null circular geodesics of the regular hairy black hole \eqref{main_metric}, with metric coefficients (\ref{hrbh}), first, the general conditions for circular motion in the given spacetime will be analyzed. The appropriate constraints for each type of geodesic -- either timelike or null -- are then imposed as special cases. Assuming that the motion of the particle is confined to the equatorial plane ($\theta = \pi/2$),  the canonical momenta associated with the cyclic coordinates are defined as:
\begin{align}
    p_t &= -f(r,\beta) \, \dot{t} \equiv -E, \\
    p_r &= \frac{\dot{r}^{2}}{f(r,\beta)}, \\
    p_{\varphi} &= r^2 \, \dot{\varphi} \equiv L,
\end{align}
where $E$ and $L$ are identified as conserved quantities corresponding to the total energy and angular momentum of the test particle, respectively. Using these definitions, the geodesic equations for $t$ and $\varphi$ are obtained:
\begin{equation}
    \dot{t} = \frac{E}{f(r,\beta)}, \quad\quad \dot{\varphi} = \frac{L}{r^2}.
\end{equation}
To determine the radial dynamics, one imposes the normalization condition for the four-velocity, $
    g_{\mu \nu} \dot{x}^\mu \dot{x}^\nu = \varepsilon,$ 
which, in the regular hairy black hole background \eqref{main_metric}, with coefficients \eqref{hrbh}, reads
\begin{equation}
    -\frac{E^2}{f(r,\beta)} + \frac{\dot{r}^2}{f(r,\beta)} + \frac{L^2}{r^2} = \varepsilon,
\end{equation}
leading to the radial geodesic equation
\begin{equation}\label{readial_eq}
    \dot{r}^2 = E^2 + \left( \varepsilon - \frac{L^2}{r^2} \right) f(r,\beta),
\end{equation}
where the parameter $\varepsilon$ distinguishes the type of geodesic: $\varepsilon = -1$ for timelike and $\varepsilon = 0$ for null geodesics. From Eq. \eqref{readial_eq}, we can identify the effective potential governing radial motion as
\begin{equation}\label{potential_eff}
    V_{\text{eff}}(r) = \left( \frac{L^2}{r^2} - \varepsilon \right) f(r,\beta). 
\end{equation}
This effective potential plays a central role in analyzing the stability of circular orbits and in computing the associated Lyapunov exponents.

\subsection{Timelike geodesics}
{Let us consider a massive test particle following a circular orbit of radius $r_c$ around the regular hairy black hole \eqref{main_metric}.} It corresponds to the case in which $\varepsilon = -1$ in Eq. \eqref{readial_eq}.

\begin{figure}[H]
    \centering
    \begin{subfigure}{0.55\textwidth}
        \centering
        \includegraphics[width=\linewidth]{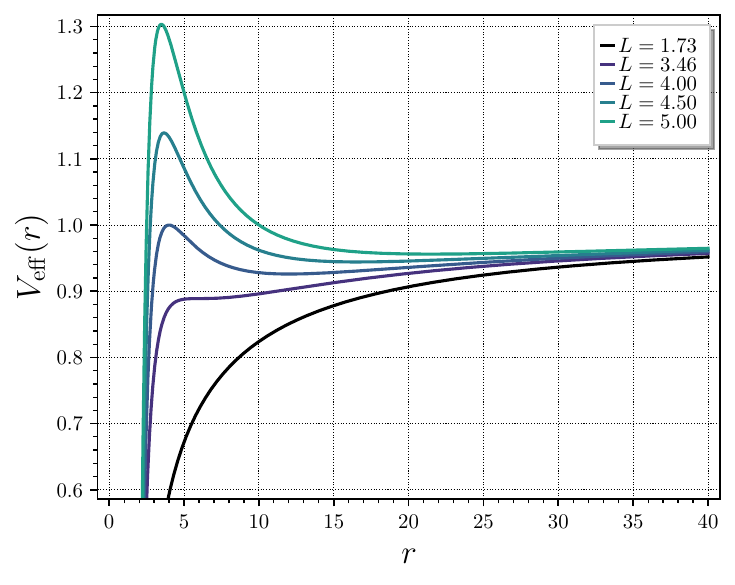}
         \caption{\footnotesize }
    \end{subfigure}%
    \begin{subfigure}{0.55\textwidth}
        \centering
        \includegraphics[width=\linewidth]{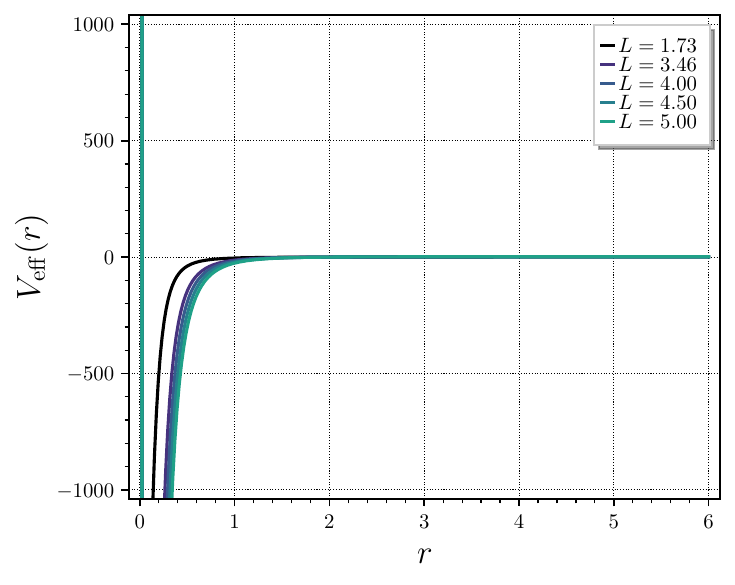}
         \caption{\footnotesize }
    \end{subfigure}

    \vspace{0.3cm}

    \begin{subfigure}{0.55\textwidth}
        \centering
        \includegraphics[width=\linewidth]{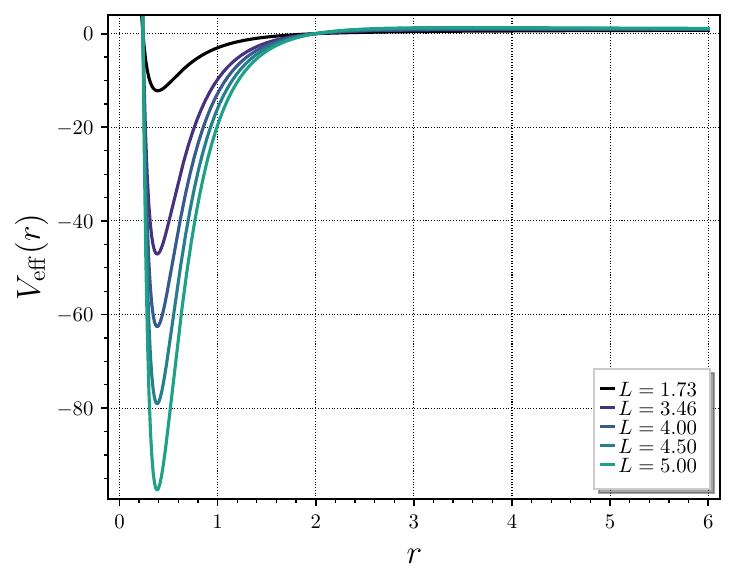}
         \caption{\footnotesize }
    \end{subfigure}%
    \begin{subfigure}{0.55\textwidth}
        \centering
        \includegraphics[width=\linewidth]{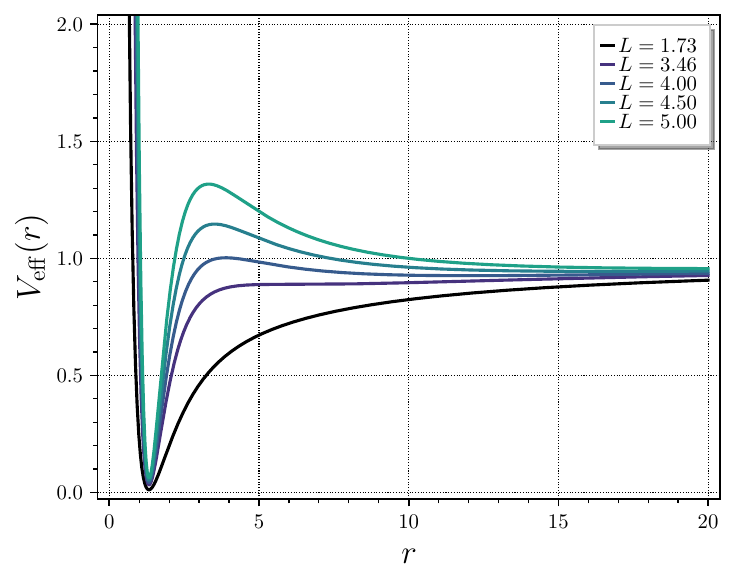}
         \caption{\footnotesize }
    \end{subfigure}

    \caption{\footnotesize  Effective potential \eqref{potential_eff} for timelike geodesics around the regular hairy black hole corresponding to: (a) Schwarzschild, (b) $\beta = 0.05$, (c) $\beta = 0.20$, and (d) $\beta = 0.39$. The radial coordinate is rescaled and expressed in units of the Schwarzschild mass.}
    \label{ptl}
\end{figure}

\noindent {Regarding Fig. \ref{ptl}, it is worth emphasizing that, in Schwarzschild spacetime, circular orbits do not exist for radii less than or equal to $r_c=3$. For $r < r_c=3$,} the lower limit for circular orbits, even unstable ones, is reached. Any geodesic, whether timelike or null, inevitably falls into the black hole. A similar behavior occurs in the case of the regular hairy black hole \eqref{main_metric}, with metric coefficients (\ref{hrbh}). We note that as the parameter $\beta$ increases, this lower limit moves progressively closer to the black hole. At these limiting values, the effective potential diverges, rendering it undefined for radii less than or equal to this threshold.

The orbital parameters can now be determined as functions of the regular hairy black hole properties and the orbital radius. For circular orbits around any spherically symmetric black hole, the conditions are sufficient. By comparing the expressions for the potential \eqref{potential_eff} and $\dot{r}^2$ \eqref{readial_eq}, we observe that the condition $\dot{r}^2 = (\dot{r}^2)' = 0$ is equivalent to $(V_{\rm{eff}})' = 0$. By studying this condition, the total energy and angular momentum for this type of orbit around the regular hairy black hole \eqref{main_metric}, with metric coefficients (\ref{hrbh}), can be  determined as 
\begin{eqnarray}
E^2 &=& \frac{2 \,  
\left[\left((r_{c} - 2)e^{\frac{r_{c}}{\beta}} + 2\right)\beta^{2} + 2 r_{c} \beta + r_{c}^{2} \right]^{2} }{
     \left[\left(2(r_{c} - 3)e^{\frac{r_{c}}{\beta}} + 6\right)\beta^{3} + 6 \, r_{c} \beta^{2}  + 3 \, r_{c}^{2} \beta + r_{c}^{3} \right] r_{c} \beta} e^{-{r_{c}}/{\beta}},\\
L^2 &=& \frac{{2 \, \beta^{3} {\left(e^{\frac{r_{c}}{\beta}} - 1\right)} - r_{c}{\left(r_{c}^{2} + r_{c} \beta + 2 \, \beta^{2}\right)} }}{2\beta^{3} \, e^{\frac{r_{c}}{\beta}}\,{\left(r_{c}  - 3 \,  + 3\, e^{-\frac{r_{c}}{\beta}}\right)}  + r_c\left(r_{c}^{2} + 3 \, r_{c} \beta + 6 \, \beta^{2}\right)} r_c^2.
\end{eqnarray}

Given the total energy and angular momentum,  the angular frequency  of a massive test particle in a circular time-like orbit of radius $r_c$ around the regular hairy black hole  can be computed as:
\begin{equation}
\Omega_0 = \sqrt{\frac{2 \, \beta^{3} {\left(1-e^{-\frac{r_{c}}{\beta}} \right)} - {\left(r_{c}^{3} + r_{c}^{2} \beta + 2 \, r_{c} \beta^{2}\right)} e^{-\frac{r_{c}}{\beta}}}{2 \, r_{c}^{3} \beta^{3}}}.
\end{equation}
Once all the necessary values have been obtained, we can calculate the Lyapunov exponent for this type of orbit as follows:
\begin{eqnarray}
\lambda_p &=& \sqrt{
\frac{r_{c}^{6} + 5 \, r_{c}^{5} \beta - e^{\frac{r_{c}}{\beta}}Z
}{e^{\frac{r_{c}}{\beta}}\left(r_{c}^{6} \beta^{3}  + 3 \, r_{c}^{5} \beta^{4}  + 6 \, r_{c}^{4} \beta^{5}\right) + 2 \, X \beta^{6}}}\\
\lambda_0 &=& 
\frac{\sqrt{2}}{W r_{c}}\sqrt{r_{c}^{6} + 5 \, r_{c}^{5} \beta - e^{\frac{r_{c}}{\beta}}\,Z},
\end{eqnarray}
where 
\begin{subequations}
\begin{eqnarray}
Z &=& \left(r_{c} - 15 \, e^{-\frac{r_{c}}{\beta}} - 2  \right) r_{c}^{4} \beta^{2} - \left(3 \, r_{c} + 26 \, e^{-\frac{r_{c}}{\beta}} - 14  \right) r_{c}^{3} \beta^{3}- \left(r_{c} + 24 \, e^{-\frac{r_{c}}{\beta}} - 12 \right) r_{c}^{2} \beta^{4} \nonumber \\
&& - 2 \, \left(r_{c} + 12 \, \left(e^{-\frac{r_{c}}{\beta}} - 1\right) \right) r_{c} \beta^{5}+ 2 \, \left[ r_{c} \left(e^{\frac{r_{c}}{\beta}} - 1\right) -12\,\left( \cosh\left(\frac{r_c}{\beta}\right) + 1\right)  \right] \beta^{6},\\
W &=&  2 \,\beta^{3}\left(r_{c}  e^{\frac{r_{c}}{\beta}} - 3 \, e^{\frac{r_{c}}{\beta}} + 3\right)+ 6 \, r_{c} \beta^{2} + 3 \, r_{c}^{2} \beta  + r_{c}^{3},\\
X &=&  r_{c}^{4} e^{\frac{2 \, r_{c}}{\beta}} - 3 \,r_{c}^{3}\, e^{\frac{r_{c}}{\beta}}\, \left( e^{\frac{\, r_{c}}{\beta}} - 1 \right).
\end{eqnarray}
\end{subequations}
\begin{figure}[H]
\centering
\includegraphics[width=0.8\linewidth]{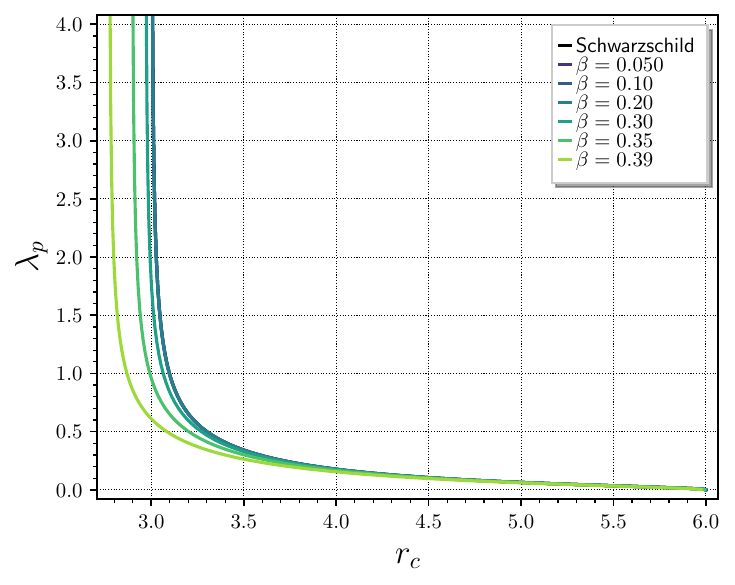}
\caption{\footnotesize Lyapunov exponents $\lambda_p$ for a circular timelike orbit around the regular hairy black hole \eqref{main_metric}. The exponents with respect to proper time diverge near $r_c = 3$ for massive orbits, indicating absolute geodesic instability and immediate growth of perturbations, which is a sign of extreme chaotic behavior. 
This behavior is important for understanding the dynamics of matter accretion onto the black hole. The radius of circular orbit $r_c$ is rescaled and expressed in units of the Schwarzschild mass.
}
\end{figure}

\begin{figure}[H]
\centering
\includegraphics[width=0.9\linewidth]{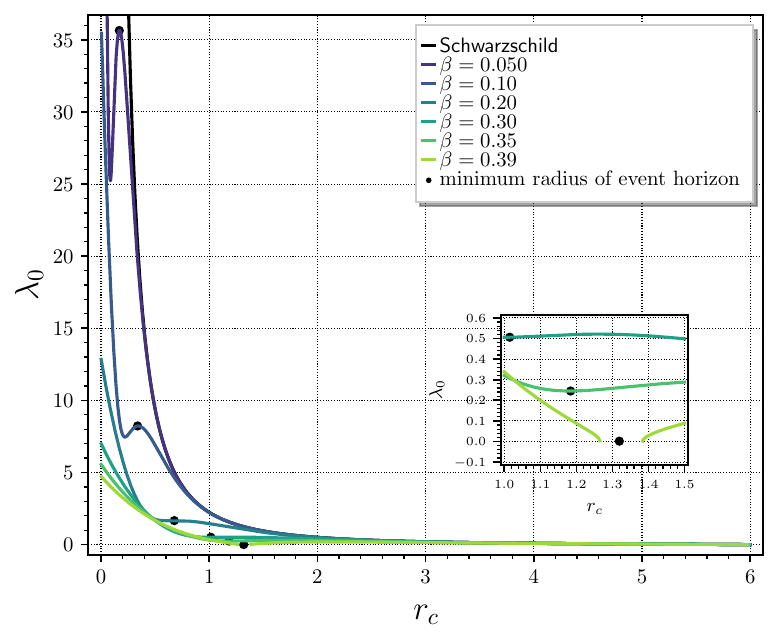}
\caption{\footnotesize Lyapunov exponents $\lambda_0$ for a circular timelike orbit around the regular hairy black hole \eqref{main_metric}. We observe that in the case of $\beta_{\rm crit} \approx 0.39$, a black hole with a single event horizon and no singularity, there is a breakdown of real values of the Lyapunov exponents near the minimal radius of the event horizon. This indicates that geodesics are unstable throughout the entire interior region, although the nature of the instability changes near the horizon. The radius of circular orbit $r_c$ is rescaled and expressed in units of the Schwarzschild mass.}
\end{figure}

The circular orbits are stable for complex values of $\lambda_0$ and unstable for real values of $\lambda_0$ \citep{Cornish:2003ig,Cardoso:2008bp}.  Table \ref{raiocritico} presents the values of critical radius, showing that for values of $r_c$ smaller than $r_{\rm crit}$, the orbit is unstable. The radius for the critical value $\beta_{\rm crit} \approx 0.39$ is also calculated.  Higher values of the hairy parameter $\beta$ indicate a modification to the stability behavior of geodesics closest to the black hole. We can see that as the parameter increases, the critical radius decreases, i.e., orbits closer to the regular hairy black hole are allowed. As expected, the critical radius is the same as the Innermost Stable Circular Orbit (ISCO), since it is the closest stable orbit to a black hole for massive particles. The inverse proportionality behavior between the hairy parameter and the ISCO radius motivates a discussion about rotating black hole mimickers in Sec. \ref{sec4}. Therein, we compare the parameter $\beta$ with the spin parameter $a$. Similarly to the hairy parameter $\beta$, in the extremal case, the larger the value of $a$, the closer the ISCO is to the black hole.

\subsection{Null geodesics}

The null circular orbits are possible when $\dot{r}^2 = 0$ at a constant radius $r = r_c$. Taking into account the regular hairy black hole metric \eqref{main_metric} and $\varepsilon=0$ in \eqref{readial_eq}, it implies that  
\begin{equation}
\dot{r}^2 = E^2 - \frac{{L}^{2}}{r^{2}} f(r,\beta),
\end{equation}
and the effective potential for massless particles reads 
\begin{equation}\label{null_potential}
    V_{\rm{eff}} = \frac{{L}^{2}}{r^{2}} f(r,\beta).
\end{equation}

\begin{figure}[H]
    \centering
    \begin{subfigure}{0.45\textwidth}
        \centering
        \includegraphics[width=\linewidth]{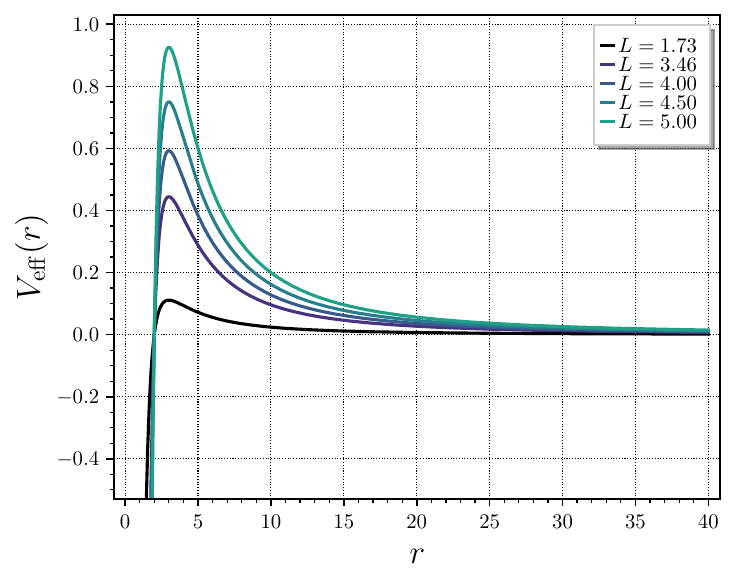}
        \caption{\footnotesize }
    \end{subfigure}%
    \begin{subfigure}{0.45\textwidth}
        \centering
        \includegraphics[width=\linewidth]{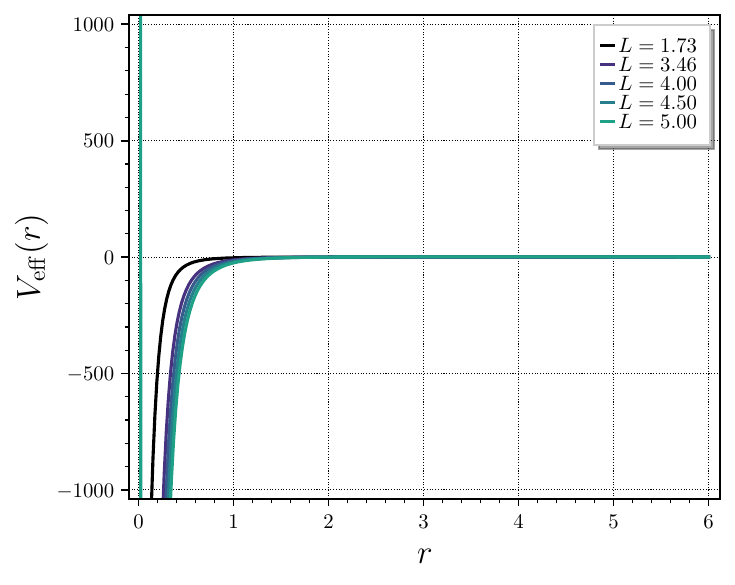}
        \caption{\footnotesize }
    \end{subfigure}

    \vspace{0.3cm}

    \begin{subfigure}{0.45\textwidth}
        \centering
        \includegraphics[width=\linewidth]{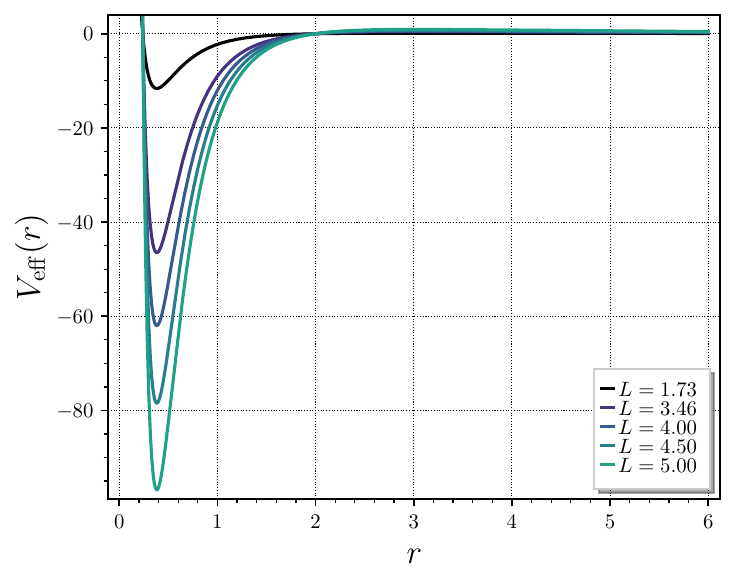}
        \caption{\footnotesize }
    \end{subfigure}%
    \begin{subfigure}{0.45\textwidth}
        \centering
        \includegraphics[width=\linewidth]{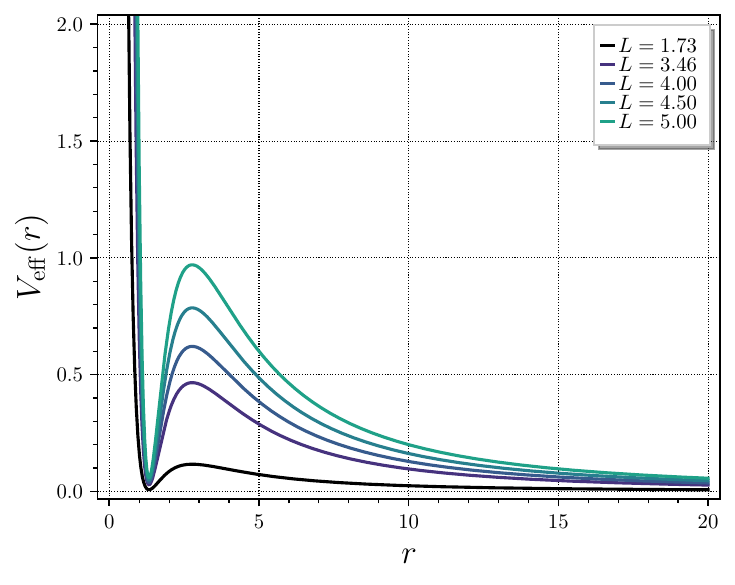}
        \caption{\footnotesize }
    \end{subfigure}

    \caption{\footnotesize Effective potential \eqref{null_potential} for timelike geodesics around the regular hairy black hole corresponding to: (a) Schwarzschild, (b) $\beta = 0.05$, (c) $\beta = 0.20$, and (d) $\beta = 0.39$. The radial coordinate is rescaled and expressed in units of the Schwarzschild mass.}
\end{figure}

\begin{figure}[H]
\centering
\includegraphics[width=0.8\linewidth]{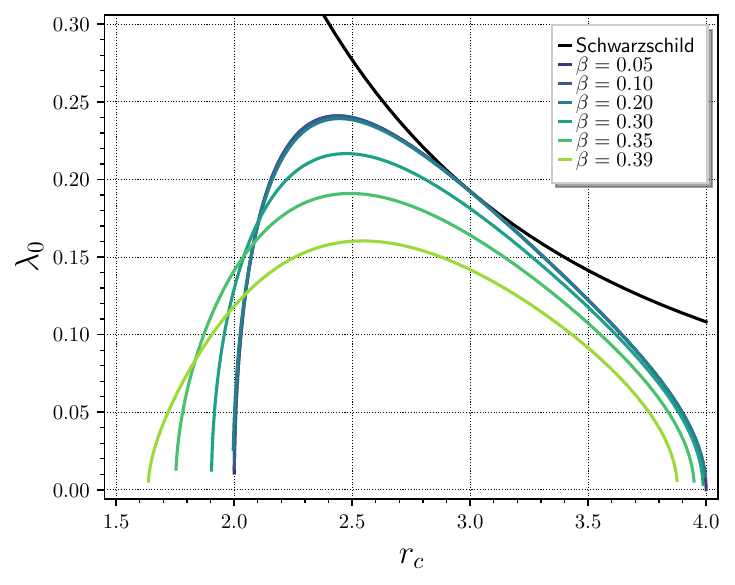}
\caption{\footnotesize Lyapunov exponents of null circular orbits. The Lyapunov exponent for this type of orbit is inversely proportional to the parameter $\beta$. However, as the value of $\beta$ increases, the concavity of the curve described by the values of $\lambda_0$ also increases. We observe that increasing $\beta$ allows stable null orbits to exist closer to the black hole. Unlike the Schwarzschild case, the regular black hole permits real positive values for the Lyapunov exponent close to $r_c =0$. These geodesics influence the shape and size of the black hole shadow, as the black hole allows a wider instability interval increasingly closer to the horizon. It is precisely the unstable photon orbits that either fall into the black hole or escape to a distant observer, generating the black hole images. The radius of circular orbit $r_c$ is rescaled and expressed in units of the Schwarzschild mass.}
\label{Lyapunov nulo}
\end{figure}

We can also derive the impact parameter $D_c$ defined as the  angular momentum-to-particle energy ratio, as 
\begin{equation}
    D_c =  \sqrt{\frac{e^{\frac{r_{c}}{ \beta}} r_{c}^{3} \beta^{2}}{{\left(r_{c} + 2 \, e^{-\frac{r_{c}}{\beta}} - 2\right)} \beta^{2} e^{\frac{r_{c}}{\beta}} + 2 \, r_{c} \beta + r_{c}^{2} }  }.
\end{equation}
Setting $\dot{r}^2 = 0$ yields the photon sphere radius of the regular hairy black hole \eqref{main_metric}. 
 Table \ref{photonsphre} presents the values of the photon sphere radius for many values of the hairy parameter $\beta$. Just as in the case of massive particle orbits, as the parameter $\beta$ increases, the radius of the photon sphere decreases, bringing it closer to the regular hairy black hole. The angular frequency $\Omega_c$ of massless particles around the regular hairy black hole reads 
\begin{equation}
\Omega_c = \sqrt{ \frac{{\left(r_{c} + 2 \, e^{-\frac{r_{c}}{\beta}} - 2\right)} \beta^{2} e^{\frac{r_{c}}{\beta}} + 2 \, r_{c} \beta + r_{c}^{2}}{r_c^3 \beta^2 e^{\frac{r_{c}}{\beta}}} } .
\end{equation}
Once all the necessary values have been found, we can determine the Lyapunov exponent for this type of orbit as follows:
\begin{equation}
\lambda_0 =  \frac{\sqrt{2}}{2{r_c^2 \beta^3}}\sqrt{{6 \, A_0\beta^{6} + 12 \, B_0 \beta^{5} + 6 \, C_0 \beta^{4} + 4 \, D_0 \beta^{3} + E_0 \beta^{2}  + 6 \, r_{c}^{5} \beta + r_{c}^{6}}},
\end{equation}
where the respective coefficients are given by 
\begin{subequations}
\begin{eqnarray}
A_0 &=& r_{c}^{2} e^{\frac{2 \, r_{c}}{\beta}} - 6 \, r_{c} e^{\frac{2 \, r_{c}}{\beta}} + 6 \, r_{c} e^{\frac{r_{c}}{\beta}} + 8 \, e^{\frac{2 \, r_{c}}{\beta}} - 16 \, e^{\frac{r_{c}}{\beta}} + 8,\\
B_0 &=& 3 \, r_{c}^{2} e^{\frac{r_{c}}{\beta}} - 8 \, r_{c} e^{\frac{r_{c}}{\beta}} + 8 \, r_{c},\\
C_0 &=& 3 \, r_{c}^{3} e^{\frac{r_{c}}{\beta}} - 8 \, r_{c}^{2} e^{\frac{r_{c}}{\beta}} + 16 \, r_{c}^{2},\\
D_0 &=& r_{c}^{4} e^{\frac{r_{c}}{\beta}} - 2 \, r_{c}^{3} e^{\frac{r_{c}}{\beta}} + 14 \, r_{c}^{3},\\
E_0 &=& r_{c}^{5} e^{\frac{r_{c}}{\beta}} - 2 \, r_{c}^{4} e^{\frac{r_{c}}{\beta}} + 22 \, r_{c}^{4}.
\end{eqnarray}
\end{subequations}
Fig. \ref{Lyapunov nulo} illustrates the Lyapunov coefficients for a circular null orbit around the regular hairy black hole \eqref{main_metric}. 
It is worth stressing that as the parameter $\beta$ increases, the peak of the graph decreases and the curve's curvature is reduced. The orbit instability range for different values of the parameter $\beta$ can also be seen in the Fig. \ref{Lyapunov nulo}. The approximation of the minimum and maximum radii, alongside the increase in the interval between them, illustrates the non-trivial dynamics of spacetime. Initially, the decrease in both radii signals a convergence towards the black hole event horizon, suggesting an enhancement in the curvature of spacetime near the black hole, influenced by the hairy parameter $\beta$. Simultaneously, the instability interval expands, shifting the instability region closer to the black hole while increasing its width. This has significant implications for the behavior of electromagnetic radiation in the vicinity of a black hole. Notably, one observable consequence is the widening of photon rings, which play a key role in shaping the black hole shadow.

\begin{table}[H]
\centering
\begin{tabular}{c c c c c c c}     
\hline\hline
$\beta$ & Schwarzschild   & 0.10    & 0.20    & 0.30    & $\beta_{\rm crit}$   \\ \hline
$r_{\rm crit}$ & 6.00000 & 5.99999 & 5.99999 & 5.99937 & 5.97673 \\ \hline
\end{tabular}
\caption{\footnotesize Critical radius for timelike circular orbits in regular hairy black hole spacetime. If $r_c < r_{\rm crit}$, the orbit is unstable. The increase of the hairy parameter $\beta$ brings the ISCO closer to the black hole event horizon.}\label{raiocritico}
\end{table}
Table \ref{raiocritico} displays the critical radius for timelike circular orbits in regular hairy black hole spacetime. 
\begin{table}[H]
\centering
\begin{tabular}{c c c c c c}     
\hline\hline
$\beta$ & Schwarzschild & 0.10    & 0.20    & 0.30    & $\beta_{\rm crit}$     \\ \hline
$r_c$ & 3.00000 & 2.99999 & 2.99936 & 2.96634 & 2.76774  \\ \hline
\end{tabular}
\caption{\footnotesize Photon sphere radius in regular hairy black hole spacetime. The increase of the hairy parameter $\beta$ brings the photon sphere closer to the black hole event horizon,  similar to the behavior observed for the ISCO.
}\label{photonsphre}
\end{table}
Table \ref{photonsphre} shows that the contracting radius of the photon sphere reveals that photon orbits are permitted closer to the black hole. This occurs due to the enhanced curvature of spacetime as the hairy parameter $\beta$ increases, further concentrating and compressing the light trajectories near the black hole.
\begin{table}[H]
\centering
\begin{tabular}{c c c c c c c}     
\hline\hline
$\beta$ & Schwarzschild   & 0.10    & 0.20    & 0.30    & 0.39     \\ \hline
$r_c^{\rm max}$ & 4.0000  & 3.99999 & 3.99993 & 3.98789 & 3.87724  \\ \hline
$r_c^{\rm min}$ & 2.00000 & 1.99999 & 1.99433 & 1.90355 & 1.64056  \\ \hline
$\Delta r$ & 2.00000 & 2.00000 & 2.00000 & 2.00560 & 2.23668 \\ \hline 
\end{tabular}
\caption{\footnotesize Range of unstability for circular null orbits. If a value belongs to the range between $r_c^{\rm max}$ and $r_c^{\rm min}$, it corresponds to an unstable orbit. We observe an increase in the minimum and maximum radii at which null orbits are allowed in the spacetime of regular black holes. As in other cases, for $\beta = 0.1$ and $\beta = 0.2$, the interval sizes do not vary within the considered precision, becoming noticeable only for larger values of $\beta$. We emphasize that, although the interval size does not always increase, the intervals always change as $\beta$ grows.
}\label{rangeinstabilide}
\end{table}
Regarding Table \ref{rangeinstabilide}, although the minimum and maximum radii decrease, signaling a convergence of the instability region toward the event horizon, the size of the region expands as $\beta$ increases. Consequently, the hairy parameter $\beta$ is inversely proportional to the distance of the interval extremes from the event horizon and directly proportional to the width of the interval.

The dynamics of timelike and null orbits present the gravity-regulating role of $\beta$. For moderate values of $\beta$, gravity becomes more intense, leading to stronger spacetime curvature. However, as $\beta \rightarrow \infty$, the curvature vanishes, resulting in the disappearance of both the photon sphere and the ISCO. A parameter that exhibits similar behavior in black holes is the rotation parameter in rotating solutions. Increasing the rotation parameter for prograde orbits causes both the ISCO and the photon sphere to move closer to the black hole, further reinforcing the comparisons between the 
$\beta$ parameter and the rotation parameter to be explored in the next section.

\subsection{Primary hairy as black hole spin mimickers}
\label{sec4}

{While remarkably successful in modeling astrophysical black holes, the Kerr solution still leaves open questions about its actual realization in nature. As hairy black holes offer promising avenues to address long-standing theoretical challenges, it is natural to ask whether such solutions remain consistent with current observational constraints. In this subsection, we test this hypothesis by comparing the ISCO of both the Kerr and the regular hairy solutions, aiming to determine whether the hair parameter can effectively mimic Kerr’s near-horizon behavior, in close analogy with the case of ultra-compact black hole mimickers~\cite{Bambi:2025wjx, Casadio:2024lgw, Yang:2023agi, Mazza:2021rgq}.}

\begin{figure}
    \centering
    \includegraphics[width=0.7\linewidth]{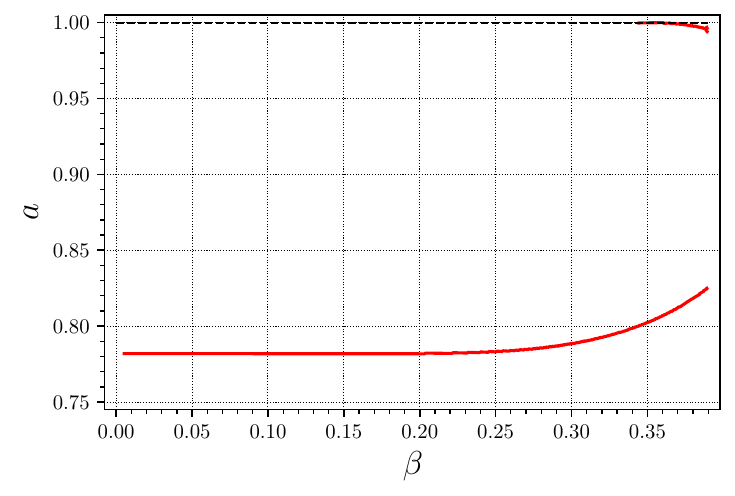}
        \caption{\footnotesize Range of the hair parameter that effectively reproduces the dynamical effects of the Kerr spin parameter $a$, as inferred from modifications in both the ISCO and the photon sphere.  Notice that for $\beta \gtrsim 0.34$, the hair parameter can mimic two distinct values of the rotation parameter. One of these corresponds to a regime approaching the Kerr extremal limit, where $a = 1$.}
    \label{fig:rotatingmimicker}
\end{figure}

The ISCO radius of the well-known Kerr solution is given by
\begin{equation}
    r_{\rm ISCO} = 3 + Z_2 - \sqrt{(3 - Z_1)(4 + Z_1 + 2Z_2)},
\end{equation}
with
\begin{eqnarray}
    Z_1 &=& 1 + \left(1-a^2\right)^{\frac{1}{3}} \left[\left(1 - a\right)^{\frac{1}{3}} + \left(1 + a\right)^{\frac{1}{3}}\right],\\
    Z_2 &=& \sqrt{3a^2 + Z_1^2}.
\end{eqnarray}
For timelike geodesics, setting  $\lambda_p = 0$ yields the ISCO radius for the regular hairy black hole. Therefore, a relationship can be established between the hairy parameter $\beta$ and the spin parameter $a$. As previously discussed, a higher hairy parameter $\beta$ results in a smaller ISCO radius. In rotating black holes, such as in the Kerr case, the ISCO radius depends on the angular momentum. Since $\beta$ decreases the ISCO radius, it may be associated with high black hole spin. As illustrated Fig.~\ref{fig:rotatingmimicker}, the hair parameter $\beta$ exhibits a behavior that closely resembles the rotation parameter $a$ in the near-extremal regime of the Kerr solution. More specifically, there exists a bounded interval, $\beta_{\min} \approx 0.00429$ and $\beta_{\max} \approx \beta_{crit}$, within which the parameter $\beta$ effectively reproduces the dynamical effects typically attributed to the Kerr spin parameter $a$, in agreement with the modifications observed in both the ISCO and the photon sphere. A similar behavior was reported for other hairy black holes in Ref. \cite{Ramos:2021jta}, where hairy black holes were shown to mimic extremal rotating black holes. {For values higher than $\beta \approx 0.34$, the parameter $\beta$ can mimic two distinct values of the Kerr spin parameter $a$, both of which correspond to regimes very close to, or even at, the extremal Kerr limit.}

\section{Quasinormal modes for the regular hairy black hole}
\label{sec3}
{Following the first detections of gravitational waves, black hole perturbation theory and the study of quasinormal modes have become highly relevant and active areas of research, offering powerful tools to probe the strong-field regime of gravity. In the context of black hole perturbation theory, an $s$-integer spin perturbation in a four-dimensional spherically symmetric black hole is governed by the Schrödinger-like equation \citep{Daghigh:2022uws,Berti:2009kk,Cardoso:2001bb}}
\begin{equation}
\label{EquacaodeRW}
\frac{d\psi_s}{dr_*^2} + \left[ \omega^2 + V_s(r) \right]\psi_s = 0,
\end{equation}
where $r_*$ is the tortoise radial coordinate, $V_s(r)$ is the Schrödinger-like potential and $\omega$ is the quasinormal frequency. Overall, the potencial $V_s(r)$ is given by \citep{Avalos:2023jeh, Stashko:2024wuq, Tang:2024txx}
\begin{equation}\label{effectivepotential}
V_s(r) = f(r) \left[\frac{\ell (\ell +1)}{r^2} + \frac{(1- s^2)}{r} \dfrac{d f(r)}{dr} \right],
\end{equation}
where $\ell$ is the multipole number and $f(r)$ is the metric function. The nature of the propagating perturbation and black hole spacetime impose the following boundary conditions 
\begin{equation}
\psi_s \sim \left\{\begin{array}{ll}\left(r - r_h\right)^{-\frac{s}{2} -i\omega r_h},\ \ r \rightarrow r_h\\r^{-s -i\omega r_h} + e^{i\omega r_h}, \;\;\; r \rightarrow + \infty\end{array}\right.,
\end{equation}
where $r_h$ and $s$ are the event horizon radius and perturbation spin, respectively. It means purely ingoing waves at the event horizon and purely outgoing waves at infinity. After establishing the boundary conditions, we proceed to discuss the methods used to solve the equation and determine the quasinormal frequencies. 

\begin{figure}[H]
    \centering
    \begin{subfigure}{0.5\textwidth}
        \centering
        \includegraphics[width=\linewidth]{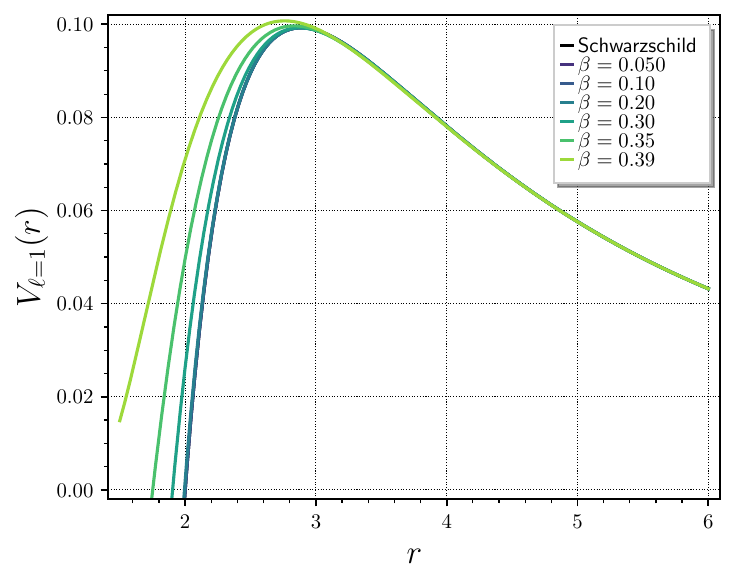}
    \end{subfigure}%
    \begin{subfigure}{0.5\textwidth}
        \centering
        \includegraphics[width=\linewidth]{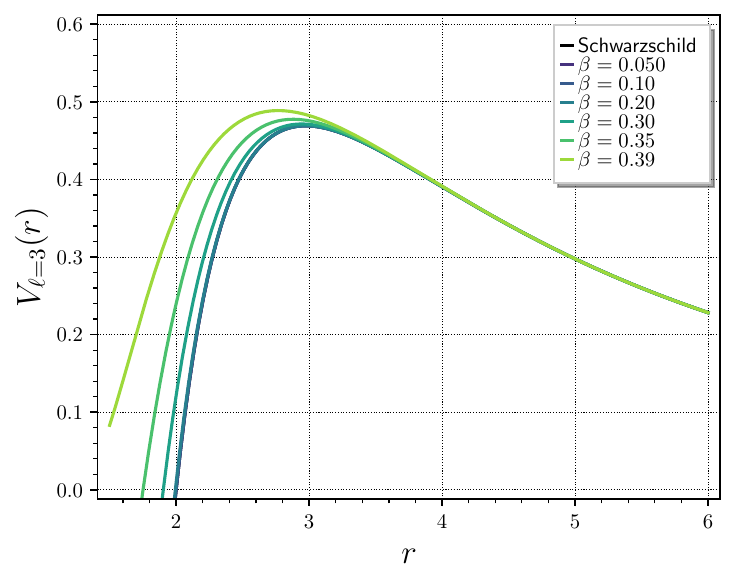}
    \end{subfigure}

    \vspace{0.3cm}

    \begin{subfigure}{0.5\textwidth}
        \centering
        \includegraphics[width=\linewidth]{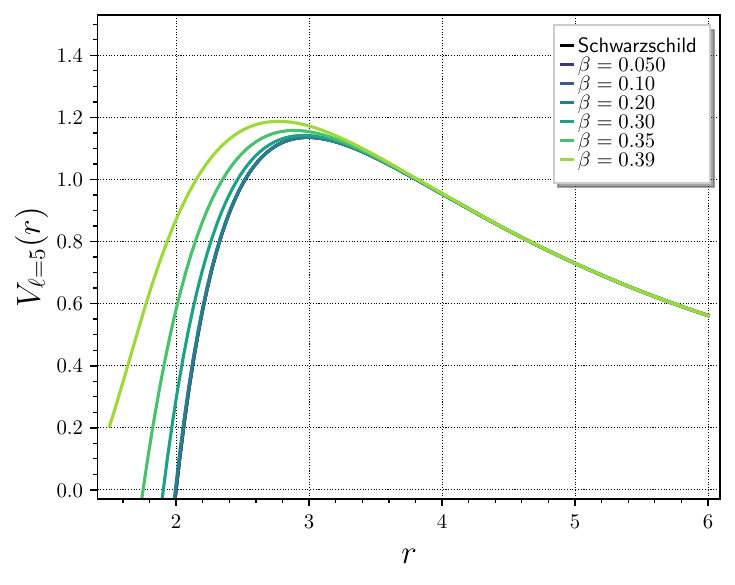}
    \end{subfigure}%
    \begin{subfigure}{0.5\textwidth}
        \centering
        \includegraphics[width=\linewidth]{V_l5_s0.pdf}
    \end{subfigure}

    \caption{\footnotesize Effective Schrödinger-like potential for scalar perturbations. As expected, the peak of the potential rises with increasing values of $\ell$. Additionally, the peak potential grows as $\beta$ increases. The radial coordinate is rescaled and expressed in units of the Schwarzschild mass.}
\label{potenciaisscalar}
\end{figure}

\begin{figure}[H]
    \centering
    \begin{subfigure}{0.5\textwidth}
        \centering
        \includegraphics[width=\linewidth]{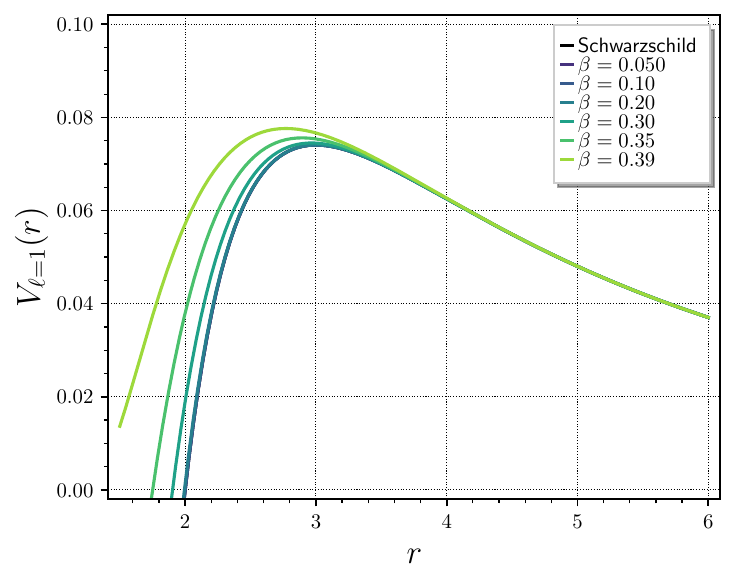}
    \end{subfigure}%
    \begin{subfigure}{0.5\textwidth}
        \centering
        \includegraphics[width=\linewidth]{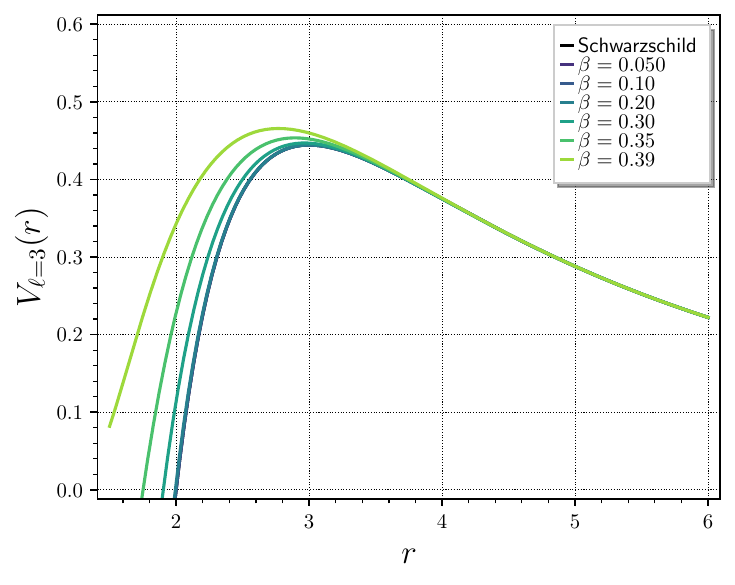}
    \end{subfigure}

    \vspace{0.3cm}

    \begin{subfigure}{0.5\textwidth}
        \centering
        \includegraphics[width=\linewidth]{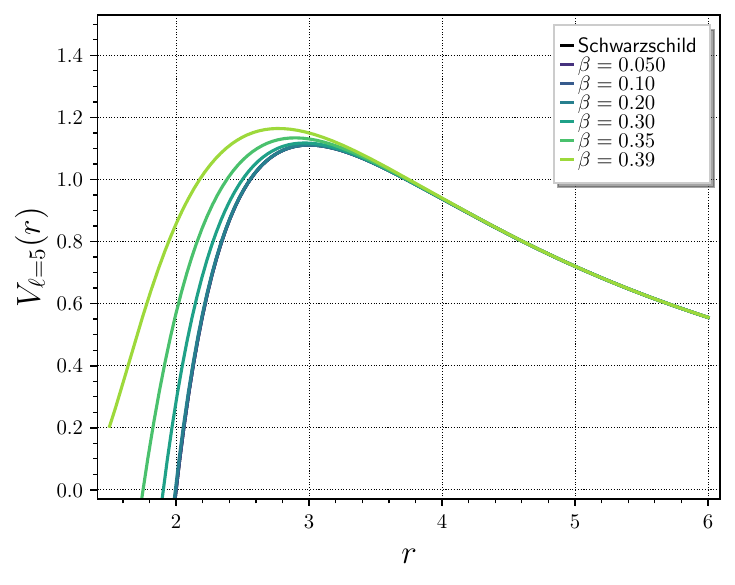}
    \end{subfigure}%
    \begin{subfigure}{0.5\textwidth}
        \centering
        \includegraphics[width=\linewidth]{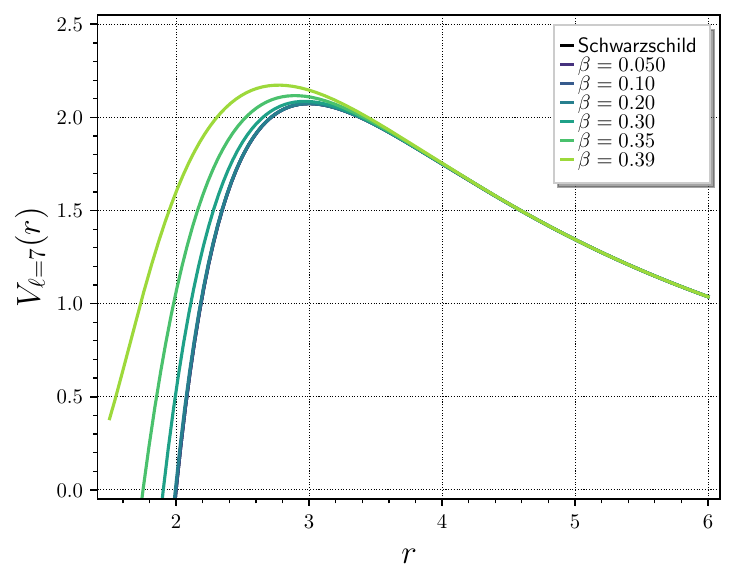}
    \end{subfigure}

    \caption{\footnotesize Effective Schrödinger-like potential for eletromagnetic perturbations. As expected, the peak of the potential rises with increasing values of $\ell$. Additionally, the peak potential grows as $\beta$ increases. The radial coordinate is rescaled and expressed in units of the Schwarzschild mass.}
    \label{potenciaiseletro}
\end{figure}

\begin{figure}[H]
    \centering
    \begin{subfigure}{0.49\textwidth}
        \centering
        \includegraphics[width=\linewidth]{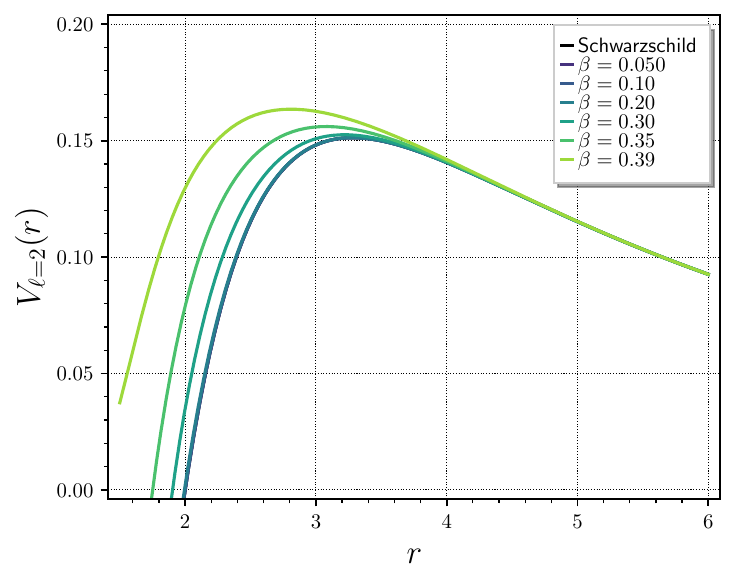}
    \end{subfigure}%
    \begin{subfigure}{0.49\textwidth}
        \centering
        \includegraphics[width=\linewidth]{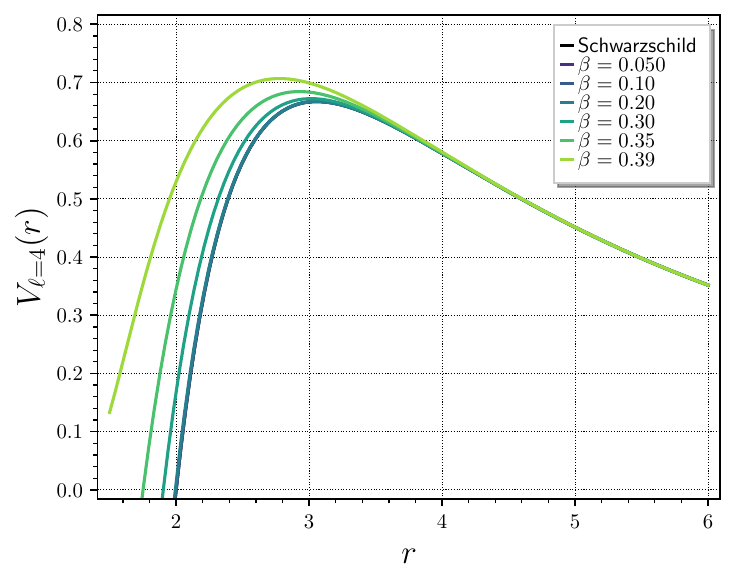}
    \end{subfigure}

    \vspace{0.3cm}

    \begin{subfigure}{0.49\textwidth}
        \centering
        \includegraphics[width=\linewidth]{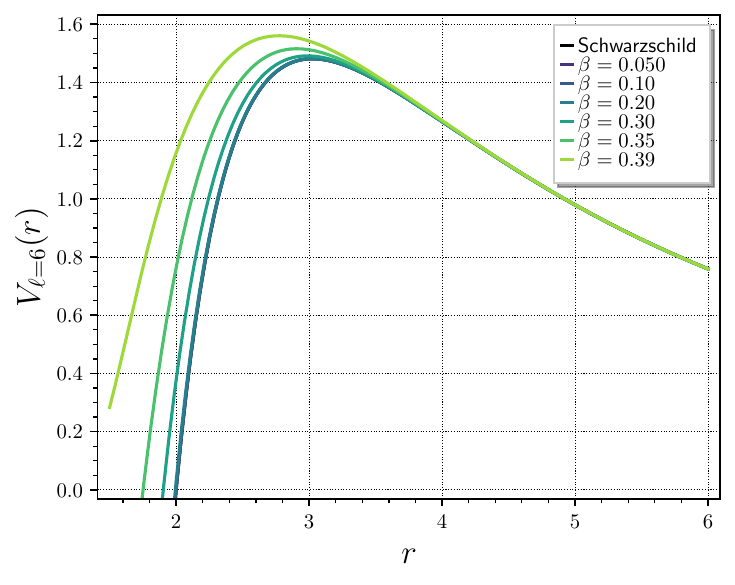}
    \end{subfigure}%
    \begin{subfigure}{0.49\textwidth}
        \centering
        \includegraphics[width=\linewidth]{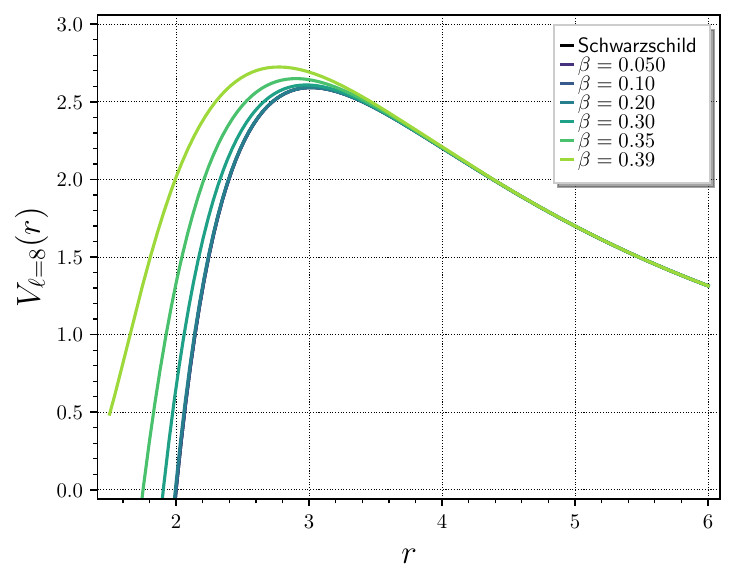}
    \end{subfigure}

\caption{\footnotesize Effective Schrödinger-like  potential for gravitational perturbations. As expected, the peak of the potential rises with increasing values of $\ell$. Additionally, the peak potential grows as $\beta$ increases. The radial coordinate is rescaled and expressed in units of the Schwarzschild mass.}
\label{potenciaisgravitacionais}
\end{figure}

{Figs.  \ref{potenciaisscalar} - \ref{potenciaisgravitacionais} show the effective Schrödinger-like potential \eqref{effectivepotential} under scalar $(s=0)$, eletromagnetic $(s=\pm 1)$ and gravitational perturbations $(s=\pm 2)$ of the regular hairy black hole, respectively. It can be seen that the larger the angular momentum $\ell$, the higher the peak of the potential. Another important feature of the potential is that the parameter $\beta$ becomes relevant only for radii approximately smaller than the ISCO.}

\subsection{WKB approximation}
{The WKB method is a semiclassical technique used to obtain approximate solutions of linear differential equations, most notably the time-independent Schrödinger equation. As previously discussed, the equation governing a spin-$s$ perturbation in a spherically symmetric spacetime takes the form of a Schrödinger-like equation, which makes the WKB technique directly applicable to black hole perturbation theory.} In this section, we use the third-order WKB approximation to derive the quasinormal frequencies of the regular hairy black hole \citep{Berti:2009kk, Konoplya:2011, Santos:2019yzk}. The quasinormal frequencies are given by \citep{Iyer:1986np,Santos:2019yzk}
\begin{align}
\omega^2 = \left[V + \sqrt{(-2V^{(2)})}\Gamma\right] - i\sqrt{\delta(-2V^{(2)})}(1 +\Omega),
\end{align}
where 
\begin{eqnarray}
 \!\!\!\!\!\!\!\!\!\!\!\!\Gamma  &=&\frac{1}{\sqrt{-2V^{(2)}}} \left[\frac{1}{8} \left(\frac{V^{(4)}}{V^{(2)}} \right)\left(\frac{1}{4} + \delta \right) - \frac{1}{288}\left(\frac{V^{(3)}}{V^{(2)}}\right)^2 (7 +60\delta) \right],\\
\!\!\!\!\!\!\!\!\!\!\!\!\Omega &=&-\frac{1}{2V^{(2)}} \left[\frac{5}{6912} \left(\frac{V^{(3)}}{V^{(2)}} \right)^2 (77 + 188\delta) - \frac{1}{384} \left(\frac{\left(V^{(3)}\right)^2 V^{(4)}}{\left(V^{(2)}\right)^3} (51 + 100\delta) \right)\right.\nonumber\\&&\left.+\frac{1}{2304}\left(\frac{V^{(4)}}{V^{(2)}} \right)^2(67+68\delta) +\frac{1}{288}\left(\frac{V^{(3)} V^{(5)}}{\left(V^{(2)}\right)^2} \right)(19 + 28\delta)- \frac{1}{288}\left(\frac{V^{(6)}}{V^{(2)}} \right)(5+4\delta)\right],
\end{eqnarray}
and
\begin{align}
\delta = \left(\frac{1}{2} + n\right)^2,
\end{align}
where $n$ denotes the overtone number. 
Here $V^{(m)}$ denotes the $m^{\rm th}$ derivatives of the Schrödinger-like effective potential,
\begin{equation}
V^{(m)} = \left.\dfrac{d^m V}{dr^m}\right|_{r=r_{\rm{max}}},
\end{equation}
evaluated at the maximum of the potential.  {We compute the quasinormal frequencies for the regular hairy black hole via 3$^{\rm rd}$-order WKB approximation. We observe that by fixing \(\ell\) and \(n\), as the hairy parameter \(\beta\) increases, the imaginary part of the frequency decreases. This decrease in the imaginary part, which is associated with the decay rate of the perturbation, suggests that the decay becomes slower as \(\beta\) increases. This may indicate a greater resistance of the system to external perturbations as \(\beta\) grows. Regarding the behavior with respect to the numbers \(\ell\) and \(n\), the description made in the case of Schwarzschild repeats. Increasing the angular momentum associated with the perturbation, i.e., \(\ell\), causes the perturbations to decay more quickly and have a higher oscillation frequency. Meanwhile, increasing the overtone number causes the oscillations to decay more rapidly. A similar behavior was described in Ref.   \cite{Cavalcanti:2022cga}.}

\begin{table}[H]
\centering
\begin{tabular}{c|c|ccccc}
\hline\hline
\multirow{2}{*}{$\ell$} & \multirow{2}{*}{$n$} & \multicolumn{5}{c}{$\omega$} \\
\cline{3-7}
 &  & Schwarzschild & $\beta = 0.10$   & $\beta = 0.20$   & $\beta = 0.30$   & $\beta = 0.39$   \\
\hline
0 & 0 & 0.1046$-$0.1152$i$ & 0.1046$-$0.1152$i$ & 0.09924$-$0.1130$i$ & 0.08485$-$0.1058$i$ & 0.08004$-$0.09957$i$ \\ \hline
\multirow{2}{*}{1} 
  & 0 & 0.2911$-$0.09800$i$ & 0.2911$-$0.09800$i$ & 0.2906$-$0.09758$i$ & 0.2883$-$0.09128$i$ & 0.2924$-$0.07785$i$ \\
  & 1 & 0.2622$-$0.3074$i$  & 0.2622$-$0.3074$i$  & 0.2593$-$0.3064$i$  & 0.2438$-$0.2907$i$  & 0.2389$-$0.2511$i$  \\ \hline
\multirow{3}{*}{2} 
  & 0 & 0.4832$-$0.09680$i$ & 0.4832$-$0.09680$i$ & 0.4830$-$0.09659$i$ & 0.4833$-$0.09211$i$ & 0.4923$-$0.07784$i$ \\
  & 1 & 0.4632$-$0.2958$i$  & 0.4632$-$0.2958$i$  & 0.4621$-$0.2951$i$  & 0.4582$-$0.2810$i$  & 0.4626$-$0.2376$i$  \\
  & 2 & 0.4317$-$0.5034$i$  & 0.4317$-$0.5034$i$  & 0.4286$-$0.5022$i$  & 0.4146$-$0.4789$i$  & 0.4078$-$0.4080$i$  \\ \hline
\multirow{4}{*}{3} 
  & 0 & 0.6752$-$0.09651$i$ & 0.6752$-$0.09651$i$ & 0.6751$-$0.09634$i$ & 0.6765$-$0.09232$i$ & 0.6898$-$0.07782$i$ \\
  & 1 & 0.6604$-$0.2923$i$  & 0.6604$-$0.2923$i$  & 0.6598$-$0.2917$i$  & 0.6593$-$0.2791$i$  & 0.6698$-$0.2352$i$  \\
  & 2 & 0.6348$-$0.4941$i$  & 0.6348$-$0.4941$i$  & 0.6330$-$0.4930$i$  & 0.6275$-$0.4709$i$  & 0.6310$-$0.3974$i$  \\
  & 3 & 0.6022$-$0.7010$i$  & 0.6022$-$0.7010$i$  & 0.5983$-$0.6993$i$  & 0.5842$-$0.6680$i$  & 0.5759$-$0.5665$i$  \\ \hline
\end{tabular}
\caption{\footnotesize Quasinormal frequencies of the regular hairy black hole under a scalar perturbation for several values of $\beta$ via the 3$^{\rm rd}$-order WKB approximation.}
\label{tab:qnm_scalar_hairywkb_combined}
\end{table}

\begin{table}[H]
\centering
\begin{tabular}{c|c|ccccc}
\hline\hline
\multirow{2}{*}{$\ell$} & \multirow{2}{*}{$n$} & \multicolumn{5}{c}{$\omega$} \\
\cline{3-7}
 &  & $\beta = 0$ & $\beta = 0.10$ & $\beta = 0.20$ & $\beta = 0.30$ & $\beta = 0.39$ \\
\hline
\multirow{2}{*}{1} 
  & 0 & 0.2459$-$0.09311$i$ & 0.2459$-$0.09311$i$ & 0.2454$-$0.09258$i$ & 0.2448$-$0.08575$i$ & 0.2533$-$0.06992$i$ \\
  & 1 & 0.2113$-$0.2958$i$  & 0.2113$-$0.2958$i$  & 0.2087$-$0.2947$i$  & 0.1982$-$0.2774$i$  & 0.1952$-$0.2293$i$  \\ \hline
\multirow{3}{*}{2} 
  & 0 & 0.4571$-$0.09507$i$ & 0.4571$-$0.09507$i$ & 0.4570$-$0.09481$i$ & 0.4580$-$0.09011$i$ & 0.4692$-$0.07507$i$ \\
  & 1 & 0.4358$-$0.2910$i$  & 0.4358$-$0.2910$i$  & 0.4348$-$0.2901$i$  & 0.4327$-$0.2755$i$  & 0.4397$-$0.2293$i$  \\
  & 2 & 0.4023$-$0.4959$i$  & 0.4023$-$0.4959$i$  & 0.3993$-$0.4944$i$  & 0.3897$-$0.4702$i$  & 0.3855$-$0.3941$i$  \\ \hline
\multirow{4}{*}{3} 
  & 0 & 0.6567$-$0.09563$i$ & 0.6567$-$0.09563$i$ & 0.6566$-$0.09544$i$ & 0.6584$-$0.09128$i$ & 0.6732$-$0.07638$i$ \\
  & 1 & 0.6415$-$0.2898$i$  & 0.6415$-$0.2898$i$  & 0.6409$-$0.2891$i$  & 0.6412$-$0.2761$i$  & 0.6538$-$0.2293$i$  \\
  & 2 & 0.6151$-$0.4901$i$  & 0.6151$-$0.4901$i$  & 0.6133$-$0.4888$i$  & 0.6096$-$0.4662$i$  & 0.6151$-$0.3902$i$  \\
  & 3 & 0.5814$-$0.6955$i$  & 0.5814$-$0.6955$i$  & 0.5775$-$0.6937$i$  & 0.5667$-$0.6616$i$  & 0.5608$-$0.5564$i$  \\ \hline
\end{tabular}
\caption{\footnotesize Quasinormal frequencies of the regular hairy black hole under electromagnetic perturbations (\(s=1\)) for several values of $\beta$ via the 3$^{\rm rd}$-order WKB approximation.}
\label{tab:qnm_vector_hairywkb_combined}
\end{table}

\begin{table}[H]
\centering
\begin{tabular}{c|c|ccccc}
\hline\hline
\multirow{2}{*}{$\ell$} & \multirow{2}{*}{$n$} & \multicolumn{5}{c}{$\omega$} \\
\cline{3-7}
 &  & Schwarzschild & $\beta = 0.10$ & $\beta = 0.20$ & $\beta = 0.30$ & $\beta = 0.39$ \\
\hline
\multirow{3}{*}{2} 
  & 0 & 0.3732$-$0.08922$i$ & 0.3732$-$0.08922$i$ & 0.3728$-$0.08868$i$ & 0.3742$-$0.08068$i$ & 0.3999$-$0.07094$i$ \\
  & 1 & 0.3460$-$0.2749$i$  & 0.3460$-$0.2749$i$  & 0.3435$-$0.2730$i$  & 0.3407$-$0.2475$i$  & 0.4169$-$0.2328$i$  \\
  & 2 & 0.3029$-$0.4711$i$  & 0.3029$-$0.4711$i$  & 0.2955$-$0.4681$i$  & 0.2811$-$0.4261$i$  & 0.4772$-$0.4226$i$  \\ \hline
\multirow{4}{*}{3} 
  & 0 & 0.5993$-$0.09273$i$ & 0.5993$-$0.09273$i$ & 0.5992$-$0.09245$i$ & 0.6019$-$0.08766$i$ & 0.6214$-$0.07130$i$ \\
  & 1 & 0.5824$-$0.2814$i$  & 0.5824$-$0.2814$i$  & 0.5816$-$0.2804$i$  & 0.5848$-$0.2658$i$  & 0.6054$-$0.2165$i$  \\
  & 2 & 0.5532$-$0.4767$i$  & 0.5532$-$0.4767$i$  & 0.5507$-$0.4747$i$  & 0.5546$-$0.4501$i$  & 0.5767$-$0.3684$i$  \\
  & 3 & 0.5157$-$0.6774$i$  & 0.5157$-$0.6774$i$  & 0.5104$-$0.6746$i$  & 0.5149$-$0.6401$i$  & 0.5395$-$0.5277$i$  \\ \hline
\end{tabular}
\caption{\footnotesize Quasinormal frequencies of the regular hairy black hole under gravitational perturbations (\(s=2\)) for several values of $\beta$ via the 3$^{\rm rd}$-order WKB approximation.}
\label{tab:qnm_tensor_hairywkb_full}
\end{table}

{Tables \ref{tab:qnm_scalar_hairywkb_combined} - \ref{tab:qnm_tensor_hairywkb_full} show the numerical results for both gravitational and scalar perturbations of the quasinormal mode frequencies of a regular hairy black hole, for several values of the parameter $\beta$ and angular momentum $\ell$. For both types of perturbations, the imaginary part of the frequency is always negative, indicating the stability of the black hole under such perturbations. In order to identify signatures of the considered hairy black hole, we compare the quasinormal frequencies obtained with the quasinormal frequencies of the Schwarzschild black hole. Once again, we observe the increase in the black hole's resistance to external perturbations as \(\beta\) increases. In general, the increase of \(\beta\) indicates a stabilization of the system and a reduction in the decay of perturbations, which characterizes the typical behavior of hairy black holes \cite{Cavalcanti:2022cga}.}

\subsection{Quasinormal modes in the eikonal limit}

{Within the framework of the eikonal limit $(\ell \gg 1)$, wave propagation is studied under the assumption that the perturbation wavelength is negligible compared to the characteristic length scales of the system \cite{Konoplya:2019hlu}.  This approximation is particularly powerful for analyzing the dynamics of gravitational waves in the strong-field region near black holes, where the gravitational field reaches extreme intensities. It has led quasinormal modes in the eikonal limit to attract considerable attention in recent years. In asymptotically flat black hole spacetimes, these modes can be interpreted in terms of unstable null circular orbits \cite{Cardoso:2008bp}. Beyond this dynamical correspondence, as explored in the present work, they may also indicate the onset of nonlinear effects and the breakdown of the linear approximation, a phenomenon known as eikonal instability  \cite{Cardoso:2008bp}. Moreover, in this regime, the effective potential becomes largely spin-independent \cite{Churilova:2019jqx}, and the quasinormal frequencies are fully determined by the parameters of the circular null geodesics, taking the form \citep{Cardoso:2008bp, Giri:2022zhf, Konoplya:2023moy}}
\begin{equation}
\omega = \ell \Omega_c - i \left(n +\frac{1}{2}\right)|\lambda_0|.
\end{equation}
Table \ref{eikonal} presents the fundamental quasinormal frequencies for many values of $\beta$.

\begin{table}[H]
\centering
\begin{tabular}{c|cccc}     
\hline\hline
\multirow{2}{*}{$\beta$} & \multicolumn{4}{c|}{$\ell$} \\ \cline{2-5} 
                   & 1                          & 2                          & 3                          & 4                          \\ \hline
Schwarzschild      & 0.17889 $-$ 0.12000$i$  & 0.35777 $-$ 0.12000$i$  & 0.53666 $-$ 0.12000$i$  & 0.71554 $-$ 0.12000$i$  \\ \hline
0.10               & 0.17889 $-$ 0.12000$i$  & 0.35777 $-$ 0.12000$i$  & 0.53666 $-$ 0.12000$i$  & 0.71554 $-$ 0.12000$i$  \\ \hline
0.20               & 0.17901 $-$ 0.11922$i$  & 0.35802 $-$ 0.11922$i$  & 0.53702 $-$ 0.11922$i$  & 0.71603 $-$ 0.11922$i$  \\ \hline
0.30               & 0.18263 $-$ 0.10828$i$  & 0.36527 $-$ 0.10828$i$  & 0.54790 $-$ 0.10828$i$  & 0.73054 $-$ 0.10828$i$  \\ \hline
0.39               & 0.19476 $-$ 0.079806$i$ & 0.38951 $-$ 0.079806$i$ & 0.58426 $-$ 0.079806$i$ & 0.77902 $-$ 0.079806$i$ \\ \hline
\end{tabular}
\caption{\footnotesize \footnotesize Fundamental quasinormal frequencies of a regular hairy black hole with a scalar field in the eikonal limit.}
\label{eikonal}
\end{table}

\section{Thermodynamic characterization of the regular hairy black hole }
\label{sec5}
 The study of black hole thermodynamics is important to ensure physical viability from GR solutions, together with studies of dynamical stability through quasinormal modes, as implemented in the previous section. Furthermore, investigating the thermodynamic properties of black holes may allow us to shed new light and gain insights into aspects of the quantum nature of gravity, which is an open puzzle in contemporary physics \cite{carlip2014black}. In this section, we will explore the thermodynamic properties of the regular hairy black hole \eqref{main_metric}, with metric coefficients \eqref{hrbh}, in both the Bekenstein-Hawking and Rényi frameworks.

The regular hairy black hole (\ref{main_metric}, \ref{hrbh}) admits a Killing vector $\upxi^{\mu} = \partial_{t}$, which is timelike for $r>r_{h}$ and null when $r = r_{h}$. 
Since the regular hairy black hole event horizon is a Killing horizon, one can calculate its surface gravity by \cite{wald2001thermodynamics}
\begin{align}
	\kappa = \lim_{r\to r_h}\sqrt{-\frac{1}{2}\nabla_{\mu}\upxi_{\nu}\nabla^{\mu}\upxi^{\nu}} = \frac{1}{2}\lim_{r\to r_h}\frac{df(r)}{dr}.
\end{align} 
Essentially, the surface gravity measures the gravitational strength at the horizon, measured by an observer at infinity. The formal relation between the surface gravity and the Hawking temperature reads 
\begin{align}
	T_{H} = \frac{\kappa}{2\pi}=\frac{1}{4\pi} \lim_{r\to r_h}\frac{df(r)}{dr}.
\end{align}
yielding the temperature at which a regular hairy black hole radiates due to quantum effects near its event horizon. 
Therefore, the Hawking temperature of the regular hairy black hole \eqref{main_metric} is given by
\begin{align}\label{Hawking_temperature}
	T_{H} = {\left(2 \,  \beta^{3} e^{\frac{{r_{h}}}{ \beta}} - 2 \, \beta^{3} - 2 \, {r_{h}}  \beta^{2} - r_{h}^{2}  \beta - r_{h}^{3}\right)} \frac{e^{-\frac{{r_{h}}}{ \beta}}}{4 \, \pi r_{h}^{2}  \beta^{3}}.
\end{align}
Eq. (\ref{Hawking_temperature}) carries the temperature associated with the quantum tensor vacuum itself, in the regular hairy black hole background. 
In the limit $\beta \rightarrow \infty$, the value $T_{H}=0$ can be recovered, agreeing with the results expected for Minkowski spacetime. On the other hand, in the limit $\beta \rightarrow 0$, the Schwarzschild black hole temperature $T_{H} = 1/4\pi r_{h}=1/8\pi M$ is recovered\footnote{Reinstating the Planck constant in the limit $\beta \rightarrow 0$,  beyond natural units here used, one should note that the  nonzero Hawking temperature $T_{H} = \hbar/8\pi G M$ associated with the Schwarzschild black hole encodes quantum effects.}.
\begin{figure}[H]
	\centering
	\includegraphics[width=0.8\hsize]{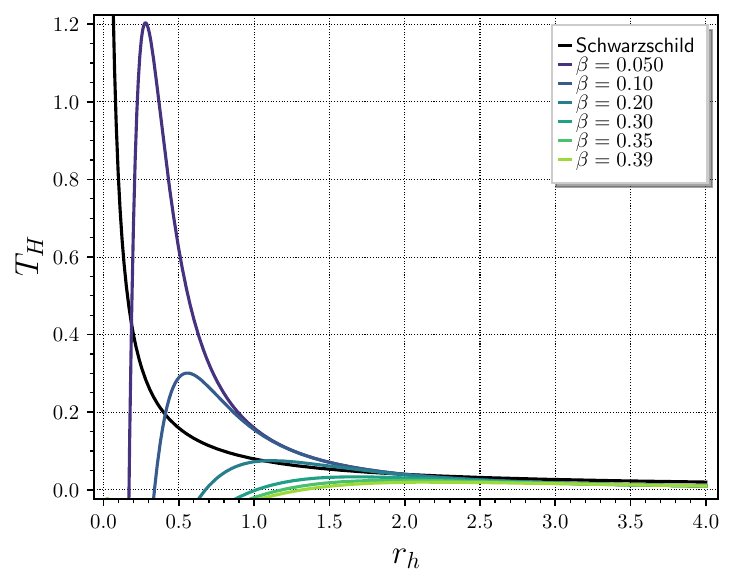}
			\caption{\footnotesize  Hawking temperature versus the horizon radius for different values of the  parameter $\beta$. The black line is the temperature of the Schwarzschild black hole, which is recovered when $\beta \rightarrow 0$. }
	\label{fig:th1}
\end{figure}
Fig. \ref{fig:th1} presents the Hawking temperature versus the horizon radius for different values of
the parameter $\beta$. The temperature equals zero when $\beta \rightarrow \infty$, whereas the Schwarzschild black hole is
recovered when $\beta \rightarrow 0$. For intermediate values of $\beta$, the temperature rises
rapidly, reaching maximum values and then decreases. Note that the higher the value of the
parameter $\beta$, the colder the black hole is, since the temperature peaks at increasingly lower
values. Another aspect one notices here is the existence of a minimum value of the horizon radius for the regular hairy black
hole to start emitting Hawking radiation, which is lacking for the Schwarzschild black
hole. Thus, it can be assumed that the hairy parameter $\beta$ imposes a minimum radius for the size of
the black hole horizon. In Table \ref{tab:rhmin}, for each value of $\beta$, the minimum
radius of the horizon ($r_{h_{{min}}}$), allowed for the black hole to exist, is displayed.

\begin{table}[H]
\centering
\renewcommand{\arraystretch}{1.2} 
\setlength{\tabcolsep}{12pt}      
\begin{tabular}{c|cccccc}
\hline\hline
\textbf{$\beta$} & 0.05 & 0.10 & 0.20 & 0.30 & 0.35 & 0.39 \\ \hline
\textbf{$r_{h_{\min}}/M$} & 0.169 & 0.338 & 0.677 & 1.015 & 1.183 & 1.319 \\
\hline\hline
\end{tabular}
\caption{\footnotesize \footnotesize Minimum horizon radius (in units of Schwarzschild mass $M$) for each value of the parameter $\beta$.}
\label{tab:rhmin}
\end{table}

Another important measure one can introduce for thermodynamic characterization of the black hole is the Bekenstein-Hawking entropy  for stationary black holes \cite{bekenstein1980black}
\begin{align}\label{sbh}
	S_{BH} = \frac{1}{4}\oint_{h} \sqrt{q}d^{2} x =\frac{A}{4},
\end{align}
where $q$ is the determinant of the induced metric on the horizon. 
The black hole entropy is proportional to the area of the horizon, where all the infalling matter accumulates according to external observers. Eq. (\ref{sbh}) reflects the number of microscopically distinct quantum states that are coarse-grained into the single macroscopic state comprising the black
hole. It also measures the number of possible microscopic states compatible with the same macroscopic parameters driving the regular hairy black hole \eqref{main_metric}.  In the case here studied, the Bekenstein-Hawking entropy reads \begin{align}\label{entropy_BH}
	S_{BH} = \pi r_{h}^{2}.
\end{align} 
With the Bekenstein-Hawking entropy, one can calculate the heat capacity of the black hole as
\begin{align}
	C_{BH} =T_{H} \frac{\partial S_{BH}}{\partial T_{H}} = T_{H} \frac{\partial S}{\partial r_{h}} \left(\frac{\partial T}{\partial r_{h}}\right)^{-1}.
\end{align}
Substituting the value of the Bekenstein-Hawking entropy given by Eq. (\ref{entropy_BH}) and using the Hawking temperature  (\ref{Hawking_temperature}) yields the following expression:
\begin{align}\label{capacidade_BH}
{{C_{BH} = -\frac{2 \pi  \, {\left(2 \, r_{h}^{2}  \beta^{4} e^{\frac{{r_{h}}}{ \beta}} - 2 \, \pi r_{h}^{2}  \beta^{4} - 2 \, \pi r_{h}^{3} \beta^{3} - \pi r_{h}^{4} \beta^{2} - \pi r_{h}^{5}  \beta\right)}}{4 \,  \beta^{4} e^{\frac{{r_{h}}}{ \beta}} - 4 \,  \beta^{4} - 4 \, {r_{h}} \beta^{3} - 2 \, r_{h}^{2} \beta^{2} - r_{h}^{4}}}}.
\end{align}
\begin{figure}[H]
	\centering
		\includegraphics[width=0.8\hsize]{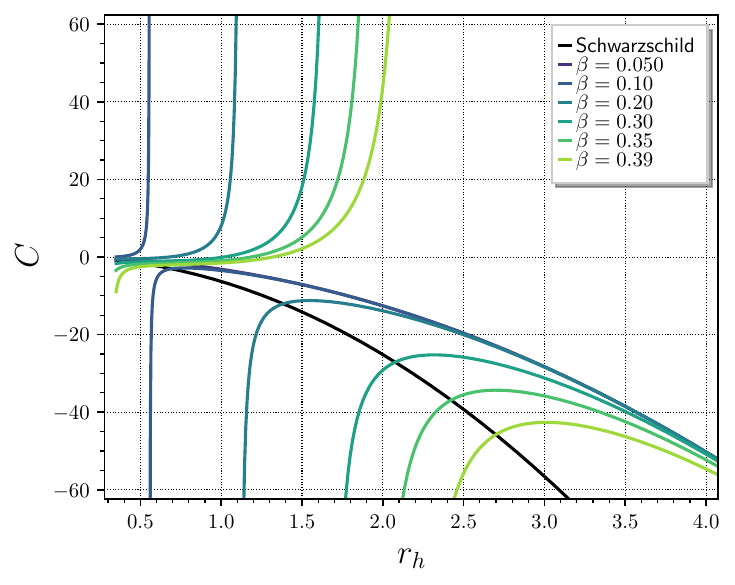}
		\caption{\footnotesize \footnotesize Heat capacity of the black hole versus the horizon radius for different values of $\beta$.}
	\label{fig:CapBH2}
\end{figure}

The heat capacity is useful for examining the local stability of a thermodynamic system. Positive
values of the heat capacity ($C_{BH}>0$) indicate that the system is stable, whereas negative values
($C_{BH}<0$) indicate that the system is unstable. However, analyzing the sign of
Eq. (\ref{capacidade_BH}) can consist of an intricate task. 
To do so, we can take Fig. \ref{fig:CapBH2} into account. 
One can see that the heat capacity has a discontinuity, which occurs at the horizon radius
value $r_{T_{max}}$ where the black hole temperature changes the sign of its derivative, reaching its maximum value. For values of the radius smaller [larger] than $r_{T_{max}}$, the regular hairy black hole
is stable [unstable].
Also, it can be noted that for values of the horizon radius sufficiently larger than
$r_{T_{max}}$, the heat capacity tends towards the heat capacity of the Schwarzschild black hole,
which is unstable.

The analysis of the thermodynamic stability of the regular hairy black hole presents a significant difference when compared to the Schwarzschild black hole solution. In fact, the Schwarzschild black hole solution is thermodynamically unstable in a canonical ensemble \cite{hawking1976black}, while the regular hairy black hole studied in the present work is stable for certain
values of horizon radii in a canonical ensemble. This stability can be associated with the parameter
$\beta$, which when tending to the limit $\beta \rightarrow 0$ recovers the Schwarzschild
instability. Although the nature of this parameter is not clear, its presence in the solution makes it
stable, from a thermodynamic point of view.

Since the Bekenstein-Hawking entropy is not extensive, for a three-dimensional black hole, the entropy is expected to be proportional to its volume, not to its area. One can scrutinize the thermodynamic stability of black holes using alternative entropy approaches in addition to the Bekenstein-Hawking entropy \cite{Feng:2024bsx,Casadio:2023pmh,Casadio:2022pla}. Among the possible alternatives, the Rényi entropy plays a prominent role, and it has been widely used to study black holes \cite{Promsiri:2020jga,Dey:2016pei}. 
The use of Rényi entropy has been motivated mainly due to its additivity, which is crucial for the analysis of thermodynamic stability via the Hessian of the entropy function. Several studies have applied the Rényi entropy to analyze the thermodynamic stability of black holes. Refs. \cite{Czinner:2015eyk,Czinner:2017tjq} studies the thermodynamic stability of the Schwarzschild and Kerr black hole solutions using the Rényi entropy \cite{Czinner:2017tjq}. In addition, Ref.  \cite{Abreu:2020vkc} investigated the Schwarzschild black hole and derived the equipartition theorem using the Rényi entropy, while the thermodynamics of the Schwarzschild-de Sitter black hole was addressed in the Rényi entropy setup in Ref. \cite{nakarachinda2021effective}. 

The Rényi entropy can be written as \cite{Czinner:2015eyk}
\begin{align}
	S_{R} = \frac{1}{\uplambda}\ln\left(1 + \uplambda S_{BH}\right),
\end{align}
where the parameter $\uplambda$ is associated with the non-locality of systems and can be related to the microscopic dynamics of the components of the system studied. Once its value is determined, it will be constant for that system. In the limit $\uplambda \rightarrow 0$, the Bekenstein-Hawking entropy is recovered.

The Rényi temperature of the regular hairy black hole can be calculated using the expression
\begin{align}
	T_{R} &= \left(1 + \uplambda S_{BH}\right) T_{H} \nonumber\\
	& = {\left(2 \, \beta^{3} e^{\frac{{r_{h}}}{ \beta}} - 2 \, \beta^{3} - 2 \, {r_{h}} \beta^{2} - r_{h}^{2} \beta - r_{h}^{3}\right)} {\left(\pi r_{h}^{2} \uplambda + 1\right)} \frac{e^{-\frac{{r_{h}}}{ \beta}}}{4 \, \pi r_{h}^{2} \beta^{3}}.
\end{align}

\begin{figure}[H]
	\centering
	\includegraphics[width=0.8\linewidth]{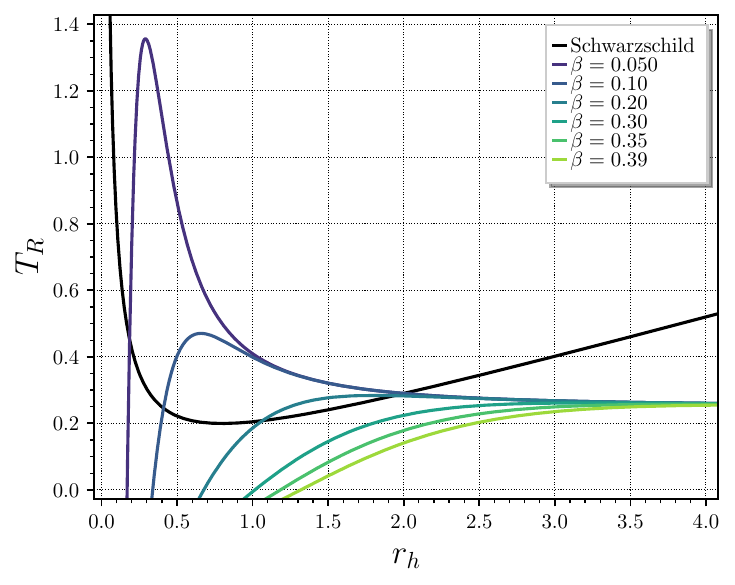}
	\caption{\footnotesize \footnotesize Rényi temperature versus radius of the horizon for different values of the $\beta$ parameter for $\uplambda=0.5$. The Rényi parameter $\lambda$ controls the deviation from standard thermodynamics.}
	\label{fig:temperaturarenyi}
\end{figure}

The Rényi temperature, displayed in Fig. \ref{fig:temperaturarenyi} for the regular hairy black hole with parameter among $ \beta=0.05$ and $\beta=0.2$, reaches a maximum value and then decreases until reaching an
asymptotic value of $T_{R}=0.25$, when $r_{h}\rightarrow \infty$. Thus, it can be observed that
the regular hairy black hole, for sufficiently large horizon radii, has the same temperature, regardless of the
value of the parameter $\beta$, when analyzed via Rényi statistics. A characteristic that remains in
the analysis of the regular hairy black hole via Rényi statistics and which is present in the Bekenstein-Hawking statistics, is the existence of a minimum radius at which the black hole starts to emit
radiation. It complies with the same values of the radius calculated in  Table \ref{tab:rhmin}.

\begin{figure}[H]
	\centering
	\includegraphics[width=0.8\linewidth]{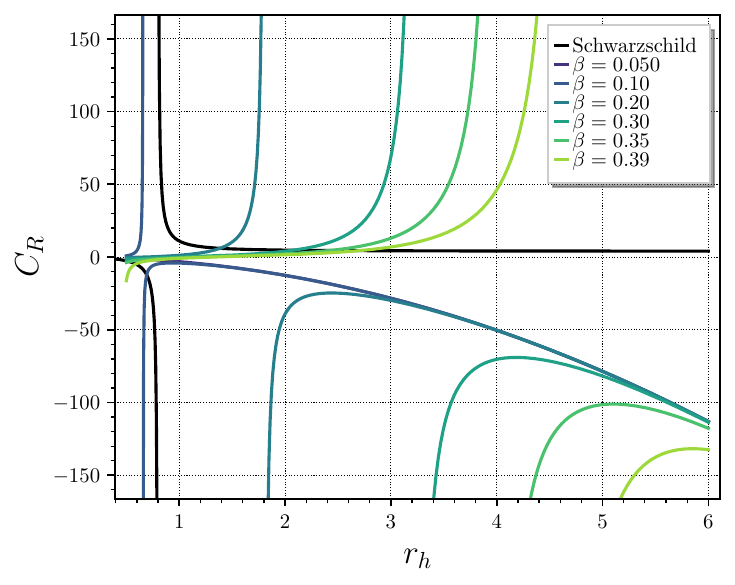}
	\caption{\footnotesize \footnotesize Rényi heat capacity of the black hole versus radius of the horizon for different values of the hairy $\beta$ parameter for $\uplambda=0.5$. The Rényi parameter $\lambda$ controls the deviation from standard thermodynamics.}
	\label{fig:capacidaderenyi}
\end{figure}

The temperature peak for the values of $\beta$ is an inflection point. The
temperature increases up to a certain value of the horizon radius $r_{h}$ and then decreases until
reaching the aforementioned asymptotic value. This inflection point is reflected in the heat capacity 
\begin{align}
C_{R} &= -\frac{2\mathcal{A}_1}{\,\mathcal{A}_2\big(1+\lambda\pi r_h^{2}\big) - 2\lambda \mathcal{A}_2\,},
\end{align}
where 
\begin{subequations}
\begin{eqnarray}\mathcal{A}_1 &=& 2\pi r_h^{2}\beta^{4}\!\left(e^{\frac{r_h}{\beta}}-1\right) - 2\pi r_h^{3}\beta^{3} - \pi r_h^{4}\beta^{2} - \pi r_h^{5}\beta, \\
\mathcal{A}_2 &=& 4\beta^{4}\!\left(e^{\frac{r_h}{\beta}}-1\right) - 4r_h\beta^{3} - 2r_h^{2}\beta^{2} - r_h^{4}. 
\end{eqnarray}
\end{subequations}
Fig. \ref{fig:capacidaderenyi} shows the thermal instability of the regular hairy black hole for
these values of the parameter $\beta$. In this case, the black hole is unstable in both Rényi and Bekenstein-Hawking statistics, as it can be seen in
Fig. \ref{fig:CapBH2}. {This behavior is quite different from the Schwarzschild black hole statistics, which are unstable for the Bekenstein-Hawking statistics but are stable within the Rényi framework.}

\section{Energy emission rate of the regular hairy black hole}

\label{sec6}
\begin{figure}[H]
	\centering
	\begin{subfigure}[b]{0.49\textwidth}
		\centering
	\includegraphics[width=\textwidth]{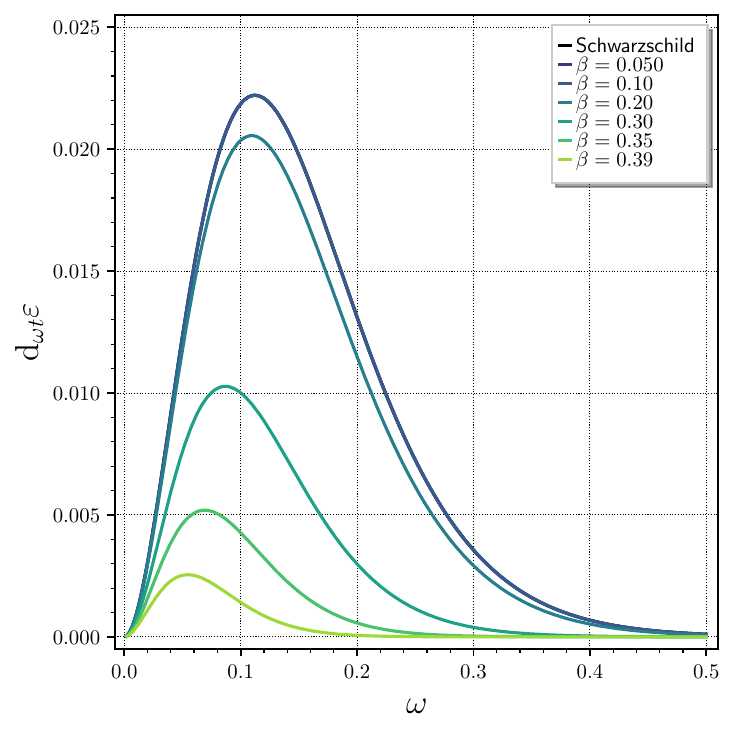}
		\caption{\footnotesize \footnotesize }
		\label{Taxa_emissao_rconst}
	\end{subfigure}
	\hfill
	\begin{subfigure}[b]{0.49\textwidth}
		\centering
	\includegraphics[width=\textwidth]{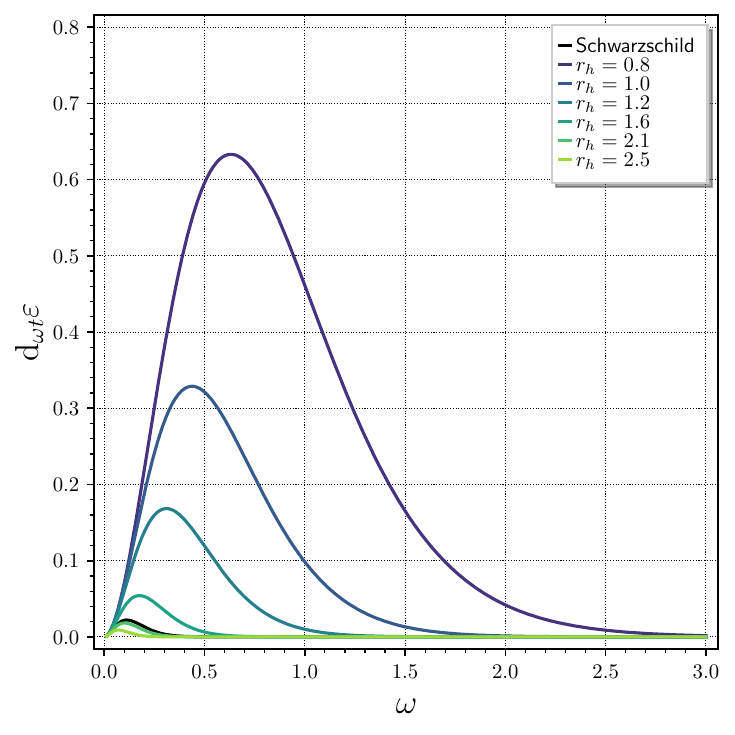}
		\caption{\footnotesize \footnotesize }
		\label{Taxa_emissao_betaconst}
	\end{subfigure}
\caption{\footnotesize \footnotesize (a) Energy emission rate versus the emission frequency $\omega$ for different values of the hairy parameter $\beta$ and $r_{h}=2$. (b) Energy emission rate versus the emission frequency $\omega$ for different values of the horizon radius and $\beta=0.1$.}
\end{figure}

Inside a black hole, beyond the event horizon, particles are constantly being created and annihilated due to quantum fluctuations \cite{ditta2023thermal}. Particles with positive energy can, by the tunneling effect, leave the inner region of the black hole for the outer region. This particle emission phenomenon comprises Hawking radiation encoding the evaporation of black holes, whose energy decreases due to the energy dissipation by particles emitted. 
The Hawking radiation, and therefore the evaporation of the black hole, is directly related to the energy emission rate, which indicates the amount of particles emitted by a black hole per unit of frequency and per unit of time is given by \cite{hawking1975particle}
\begin{align}\label{energy_rate_emisson}
	\frac{d^{2}\varepsilon}{d\omega dt} = \frac{2\pi^{2}\sigma_{lim}}{e^{\frac{\omega}{T}}-1}\omega^{3},
\end{align}
where $T$ is the temperature and $\omega$ is the emission frequency; $\sigma_{lim}$ is the area of the high-energy absorption cross section for an observer located
at infinity \cite{wei2013observing}, which varies around a limiting value given by the area of the cross-section of the black
hole event horizon \cite{mashhoon1973scattering}.
\begin{align}\label{cross_section}
	\sigma_{lim} \approx \pi r_{h}^{2}.
\end{align}
Substituting Eq. \eqref{cross_section} and the Hawking temperature \eqref{Hawking_temperature} into Eq. \eqref{energy_rate_emisson} yields  the following expression \cite{ditta2023thermodynamic,ditta2022thermal,panah2020charged,chen2024quasi}
\begin{align}\label{aa1}
	\frac{d^{2}\varepsilon}{d\omega dt} = \frac{2 \, \pi^{3} r_{h}^{2} \omega^{3}}{\exp{g(\beta,\omega,r_h)} - 1},
\end{align}
where 
\begin{eqnarray}
g(\beta,\omega,r_h)=\frac{4 \, \pi r_{h}^{2} \beta^{3} \omega e^{\frac{{r_{h}}}{ \beta}}}{2 \, \beta^{3} e^{\frac{{r_{h}}}{ \beta}} - 2 \, \beta^{3} - 2 \, {r_{h}}  \beta^{2} - r_{h}^{2} \beta - r_{h}^{3}}.
\end{eqnarray}
Eq. (\ref{aa1}) is depicted as a function of the regular hairy black hole parameter $\beta$ and $r_h$, in Fig. \ref{Taxa_emissao_betaconst}. 

In Fig. \ref{Taxa_emissao_rconst}, the energy emission rate for a constant value of
the radius of the black hole horizon is displayed, for different values of the parameter $\beta$. It is worth noting that the higher the value of the parameter $\beta$, the lower the black hole emission rate, which complies with the black hole Hawking temperature. The higher the value of $\beta$,
the lower the Hawking temperature of the black hole. We can also notice that the higher the value of
$\beta$, the lower the maximum value of the emission frequency is.  When the
energy emission rate is plotted as a function of frequency, for fixed values of the fixed parameter $\beta$, 
and vary the possible values of the radius of the event horizon, as in Fig.  \ref{Taxa_emissao_betaconst}, we notice that the smaller the black hole, the higher its emission rate
and the peak of its emission rate occurs at a higher frequency.

\section{Concluding remarks}

{The analyses performed in the present paper highlight the significant role of the hairy parameter $\beta$ in shaping the spacetime structure and dynamics around a regular hairy black hole. As $\beta$ increases, both the photon sphere and the innermost stable circular orbit (ISCO) move closer to the event horizon, reflecting an enhanced spacetime curvature near the black hole. This shift results in a wider instability interval and allows photon orbits and massive particle orbits to exist closer to the black hole, with observable consequences such as the widening of photon rings, which directly influence the black hole shadow. The behavior of $\beta$ also reveals its gravity-regulating role: moderate values increase the gravitational intensity, while in the limit $\beta \to \infty$, the curvature vanishes, causing the disappearance of the photon sphere and the ISCO. Further, the hairy $\beta$ parameter mirrors the influence of the rotation parameter in extremal Kerr black holes, where higher spin values similarly move the ISCO and photon sphere closer to the horizon. Consequently, the hairy  $\beta$ parameter can mimic the effect of the rotational parameter, particularly in near-extremal scenarios.

Perturbative analyses confirm the stability of the regular hairy black hole under both gravitational and scalar perturbations, as indicated by the negative imaginary part of quasinormal mode frequencies. Additionally, the effective potential increases with both the angular momentum $\ell$ and the parameter $\beta$, reinforcing the link between $\beta$ and stronger spacetime curvature effects.

The thermodynamic aspects of a regular hairy black hole were also explored within both the Bekenstein–Hawking and Rényi approaches. In particular, it was shown that black holes exhibit a maximum temperature for each value of the parameter $\beta$, which is reflected in a discontinuity in the heat capacity. A detailed analysis of the heat capacity revealed the existence of horizon radii for which the black hole is stable, and others for which it is unstable. That is, there are specific horizon radii that allow for the physical existence of regular hairy black holes. A distinctive feature of the solution discussed is that, in the Rényi framework, it behaves similarly to the Bekenstein–Hawking statistics, unlike the Schwarzschild solution, which is stable. The analyses of black hole energy emission rate show that the greater the value of the parameter $\beta$ and the horizon radius $r_h$, the lower the black hole's evaporation rate. It could have consequences for the expected amount of primordial black holes.

Overall, the hairy parameter $\beta$ emerges as a crucial factor controlling orbital dynamics, stability, thermodynamics, and the observable properties of a regular black hole. Under certain circumstances, it is possible to effectively replicate the characteristics of high-rotating Kerr-type spacetimes.}

\label{sec7}
\subsection*{\textbf{Acknowledgements}}
RCdP and KSA thanks to Coordenação de Aperfeiçoamento de Pessoal de Nível Superior – Brasil (CAPES) – Finance Code 001. RTC thanks the National Council for Scientific and Technological Development - CNPq (Grant No. 401567/2023-0) for partial financial support. The work of RdR is supported by The S\~ao Paulo Research Foundation (FAPESP) 
(Grants No. 2021/01089-1 and No. 2024/05676-7) and CNPq (Grants No. 303742/2023-2 and No. 401567/2023-0).

\bibliography{references}

\providecommand{\newblock}{}
\begin{thebibliography}{100}
\expandafter\ifx\csname url\endcsname\relax
  \def\url#1{{\tt #1}}\fi
\expandafter\ifx\csname urlprefix\endcsname\relax\def\urlprefix{URL }\fi
\providecommand{\eprint}[2][]{\url{#2}}

\bibitem{LIGOScientific:2016lio}
Abbott B~P {\em et~al.\/} (LIGO Scientific, Virgo) 2016 {\em Phys. Rev.
  Lett.\/} {\bf 116} 221101 [Erratum: Phys.Rev.Lett. 121, 129902 (2018)]
  (\textit{Preprint} \eprint{1602.03841})

\bibitem{LIGOScientific:2019fpa}
Abbott B~P {\em et~al.\/} (LIGO Scientific, Virgo) 2019 {\em Phys. Rev. D\/}
  {\bf 100} 104036 (\textit{Preprint} \eprint{1903.04467})

\bibitem{Ovalle:2017fgl}
Ovalle J 2017 {\em Phys. Rev. D\/} {\bf 95} 104019 (\textit{Preprint}
  \eprint{1704.05899})

\bibitem{Ovalle:2018gic}
Ovalle J 2019 {\em Phys. Lett. B\/} {\bf 788} 213--218 (\textit{Preprint}
  \eprint{1812.03000})

\bibitem{Maurya:2024zao}
Maurya S~K, Jasim M~K, Errehymy A, Boshkayev K, Mustafa G and Dayanandan B 2024
  {\em Phys. Dark Univ.\/} {\bf 46} 101665

\bibitem{LinaresCedeno:2019aul}
Linares~Cede{\~n}o F~X and Contreras E 2020 {\em Phys. Dark Univ.\/} {\bf 28}
  100543 (\textit{Preprint} \eprint{1907.04892})

\bibitem{Tello-Ortiz:2020ydf}
Tello-Ortiz F 2020 {\em Eur. Phys. J. C\/} {\bf 80} 413

\bibitem{Maurya:2025jzj}
Maurya S~K, Kiroriwal S, Kumar J and Aziz A 2025 {\em Eur. Phys. J. C\/} {\bf
  85} 456

\bibitem{Pradhan:2024hne}
Pradhan S, Bhar P, Mandal S, Sahoo P~K and Bamba K 2025 {\em Eur. Phys. J. C\/}
  {\bf 85} 127 (\textit{Preprint} \eprint{2408.03967})

\bibitem{Yousaf:2024src}
Yousaf Z, Bamba K, Almutairi B, Hashimoto Y and Khan S 2024 {\em Phys. Dark
  Univ.\/} {\bf 46} 101629 (\textit{Preprint} \eprint{2408.12132})

\bibitem{Maurya:2024ylr}
Maurya S~K, Errehymy A, Newton~Singh K, Aziz A, Hansraj S and Ray S 2024 {\em
  Astrophys. J.\/} {\bf 972} 175

\bibitem{daRocha:2020gee}
da~Rocha R and Tomaz A~A 2020 {\em Eur. Phys. J. C\/} {\bf 80} 857
  (\textit{Preprint} \eprint{2005.02980})

\bibitem{DaRocha:2019fjr}
da~Rocha R and Tomaz A~A 2019 {\em Eur. Phys. J. C\/} {\bf 79} 1035
  (\textit{Preprint} \eprint{1905.01548})

\bibitem{Casadio:2023mgl}
Casadio R and da~Rocha R 2023 {\em Eur. Phys. J. C\/} {\bf 83} 537
  (\textit{Preprint} \eprint{2305.15752})

\bibitem{Estrada:2019aeh}
Estrada M 2019 {\em Eur. Phys. J.\/} {\bf C79} 918 (\textit{Preprint}
  \eprint{1905.12129})

\bibitem{Gabbanelli:2019txr}
Gabbanelli L, Ovalle J, Sotomayor A, Stuchlik Z and Casadio R 2019 {\em Eur.
  Phys. J.\/} {\bf C79} 486 (\textit{Preprint} \eprint{1905.10162})

\bibitem{Leon:2023nbj}
Leon P and Las~Heras C 2023 {\em Eur. Phys. J. C\/} {\bf 83} 260

\bibitem{Ramos:2021drk}
Ramos A, Arias C, Fuenmayor E and Contreras E 2021 {\em Eur. Phys. J. C\/} {\bf
  81} 203 (\textit{Preprint} \eprint{2103.05039})

\bibitem{Rincon:2019jal}
Rinc\'on A, Gabbanelli L, Contreras E and Tello-Ortiz F 2019 {\em Eur. Phys. J.
  C\/} {\bf 79} 873 (\textit{Preprint} \eprint{1909.00500})

\bibitem{Panotopoulos:2018law}
Panotopoulos G and Rinc\'on A 2018 {\em Eur. Phys. J.\/} {\bf C78} 851
  (\textit{Preprint} \eprint{1810.08830})

\bibitem{Singh:2019ktp}
Singh K~N, Maurya S~K, Jasim M~K and Rahaman F 2019 {\em Eur. Phys. J. C\/}
  {\bf 79} 851

\bibitem{Ovalle:2018ans}
Ovalle J, Casadio R, da~Rocha R, Sotomayor A and Stuchlik Z 2018 {\em EPL\/}
  {\bf 124} 20004 (\textit{Preprint} \eprint{1811.08559})

\bibitem{PerezGraterol:2018eut}
P\'erez~Graterol R 2018 {\em Eur. Phys. J. Plus\/} {\bf 133} 244

\bibitem{Heras:2018cpz}
Heras C~L and Leon P 2018 {\em Fortsch. Phys.\/} {\bf 66} 1800036
  (\textit{Preprint} \eprint{1804.06874})

\bibitem{Torres:2019mee}
Torres-S\'anchez V~A and Contreras E 2019 {\em Eur. Phys. J.\/} {\bf C79} 829
  (\textit{Preprint} \eprint{1908.08194})

\bibitem{Tello-Ortiz:2021kxg}
Tello-Ortiz F, Maurya S~K and Bargue\~no P 2021 {\em Eur. Phys. J. C\/} {\bf
  81} 426

\bibitem{Zubair:2023cvu}
Zubair M, Azmat H and Jameel H 2023 {\em Eur. Phys. J. C\/} {\bf 83} 905

\bibitem{Bamba:2023wok}
Bamba K, Bhatti M~Z, Yousaf Z and Shoukat Z 2023 {\em Eur. Phys. J. C\/} {\bf
  83} 1033 (\textit{Preprint} \eprint{2307.10399})

\bibitem{Maurya:2023uiy}
Maurya S~K, Mustafa G, Ray S, Dayanandan B, Aziz A and Errehymy A 2023 {\em
  Phys. Dark Univ.\/} {\bf 42} 101284

\bibitem{Iqbal:2025xqf}
Iqbal N, Amir M, Alshammari M, Mohammed W~W and Ilyas M 2025 {\em Eur. Phys. J.
  C\/} {\bf 85} 428

\bibitem{Contreras:2021xkf}
Contreras E and Fuenmayor E 2021 {\em Phys. Rev. D\/} {\bf 103} 124065
  (\textit{Preprint} \eprint{2107.01140})

\bibitem{Sharif:2020lbt}
Sharif M and Majid A 2020 {\em Chin. J. Phys.\/} {\bf 68} 406--418

\bibitem{Meert:2020sqv}
Meert P and da~Rocha R 2021 {\em Nucl. Phys. B\/} {\bf 967} 115420
  (\textit{Preprint} \eprint{2006.02564})

\bibitem{daRocha:2014dla}
da~Rocha R and Hoff~da Silva J~M 2014 {\em EPL\/} {\bf 107} 50001
  (\textit{Preprint} \eprint{1408.2402})

\bibitem{Babichev:2013cya}
Babichev E and Charmousis C 2014 {\em JHEP\/} {\bf 08} 106 (\textit{Preprint}
  \eprint{1312.3204})

\bibitem{Sotiriou:2013qea}
Sotiriou T~P and Zhou S~Y 2014 {\em Phys. Rev. Lett.\/} {\bf 112} 251102
  (\textit{Preprint} \eprint{1312.3622})

\bibitem{Antoniou:2017acq}
Antoniou G, Bakopoulos A and Kanti P 2018 {\em Phys. Rev. Lett.\/} {\bf 120}
  131102 (\textit{Preprint} \eprint{1711.03390})

\bibitem{Hayward:2005gi}
Hayward S~A 2006 {\em Phys. Rev. Lett.\/} {\bf 96} 031103 (\textit{Preprint}
  \eprint{gr-qc/0506126})

\bibitem{Ovalle:2020kpd}
Ovalle J, Casadio R, Contreras E and Sotomayor A 2021 {\em Phys. Dark Univ.\/}
  {\bf 31} 100744 (\textit{Preprint} \eprint{2006.06735})

\bibitem{Liang:2024xif}
Liang Y, Lyu X and Tao J 2024 {\em Commun. Theor. Phys.\/} {\bf 76} 085402

\bibitem{Avalos:2023ywb}
Avalos R, Bargue\~no P and Contreras E 2023 {\em Fortsch. Phys.\/} {\bf 2023}
  2200171 (\textit{Preprint} \eprint{2303.04119})

\bibitem{Zhang:2022niv}
Zhang C~M, Zhang M and Zou D~C 2023 {\em Chin. Phys. C\/} {\bf 47} 015106
  (\textit{Preprint} \eprint{2208.06830})

\bibitem{Ditta:2023arv}
Ditta A, Javed F, Maurya S~K, Mustafa G and Atamurotov F 2023 {\em Phys. Dark
  Univ.\/} {\bf 42} 101345

\bibitem{Mahapatra:2022xea}
Mahapatra S and Banerjee I 2023 {\em Phys. Dark Univ.\/} {\bf 39} 101172
  (\textit{Preprint} \eprint{2208.05796})

\bibitem{Albalahi:2024vpy}
Albalahi A~M, Yousaf Z, Ali A and Khan S 2024 {\em Eur. Phys. J. C\/} {\bf 84}
  9

\bibitem{Misyura:2024fho}
Misyura M, Rincon A and Vertogradov V 2024 {\em Phys. Dark Univ.\/} {\bf 46}
  101717 (\textit{Preprint} \eprint{2405.05370})

\bibitem{Cavalcanti:2016mbe}
Cavalcanti R~T, da~Silva A~G and da~Rocha R 2016 {\em Class. Quant. Grav.\/}
  {\bf 33} 215007 (\textit{Preprint} \eprint{1605.01271})

\bibitem{Cavalcanti:2022cga}
Cavalcanti R~T, de~Paiva R~C and da~Rocha R 2022 {\em Eur. Phys. J. Plus\/}
  {\bf 137} 1185 (\textit{Preprint} \eprint{2203.08740})

\bibitem{Yang:2022ifo}
Yang Y, Liu D, \"Ovg\"un A, Long Z~W and Xu Z 2023 {\em Phys. Rev. D\/} {\bf
  107} 064042 (\textit{Preprint} \eprint{2203.11551})

\bibitem{Li:2022hkq}
Li Z 2023 {\em Phys. Lett. B\/} {\bf 841} 137902 (\textit{Preprint}
  \eprint{2212.08112})

\bibitem{Cavalcanti:2022adb}
Cavalcanti R~T, Alves K~d~S and Hoff~da Silva J~M 2022 {\em Universe\/} {\bf 8}
  363 (\textit{Preprint} \eprint{2207.03995})

\bibitem{Guimaraes:2025jsh}
Guimar{\~a}es V~F, Cavalcanti R~T and da~Rocha R 2025  (\textit{Preprint}
  \eprint{2506.20044})

\bibitem{Avalos:2023jeh}
Avalos R and Contreras E 2023 {\em Eur. Phys. J. C\/} {\bf 83} 155
  (\textit{Preprint} \eprint{2302.09148})

\bibitem{Tello-Ortiz:2024mqg}
Tello-Ortiz F, Avalos R, G\'omez-Leyton Y and Contreras E 2024 {\em Phys. Dark
  Univ.\/} {\bf 46} 101547

\bibitem{Meng:2025glf}
Meng K, Zhao J, Deng M, Li C and Yang N 2025 {\em Phys. Lett. B\/} {\bf 868}
  139630

\bibitem{daRocha:2017cxu}
da~Rocha R 2017 {\em Phys. Rev. D\/} {\bf 95} 124017 (\textit{Preprint}
  \eprint{1701.00761})

\bibitem{Casadio:2017sze}
Casadio R, Nicolini P and da~Rocha R 2018 {\em Class. Quant. Grav.\/} {\bf 35}
  185001 (\textit{Preprint} \eprint{1709.09704})

\bibitem{Poisson:1990eh}
Poisson E and Israel W 1990 {\em Phys. Rev. D\/} {\bf 41} 1796--1809

\bibitem{Bonanno:2022jjp}
Bonanno A, Khosravi A~P and Saueressig F 2023 {\em Phys. Rev. D\/} {\bf 107}
  024005 (\textit{Preprint} \eprint{2209.10612})

\bibitem{Carballo-Rubio:2022kad}
Carballo-Rubio R, Di~Filippo F, Liberati S, Pacilio C and Visser M 2022 {\em
  JHEP\/} {\bf 09} 118 (\textit{Preprint} \eprint{2205.13556})

\bibitem{Tang:2024txx}
Tang C, Ling Y, Jiang Q~Q and Li G~P 2024 {\em Eur. Phys. J. C\/} {\bf 84} 1296
  (\textit{Preprint} \eprint{2411.01764})

\bibitem{Stashko:2024wuq}
Stashko O 2024 {\em Phys. Rev. D\/} {\bf 110} 084016 (\textit{Preprint}
  \eprint{2407.07892})

\bibitem{Konoplya:2024hfg}
Konoplya R~A and Zhidenko A 2024 {\em Phys. Rev. D\/} {\bf 109} 104005
  (\textit{Preprint} \eprint{2403.07848})

\bibitem{Balart:2023odm}
Balart L, Panotopoulos G and Rinc\'on A 2023 {\em Fortsch. Phys.\/} {\bf 71}
  2300075 (\textit{Preprint} \eprint{2309.01910})

\bibitem{Ovalle:2023ref}
Ovalle J, Casadio R and Giusti A 2023 {\em Phys. Lett. B\/} {\bf 844} 138085
  (\textit{Preprint} \eprint{2304.03263})

\bibitem{EventHorizonTelescope:2022wkp}
Akiyama K {\em et~al.\/} (Event Horizon Telescope) 2022 {\em Astrophys. J.
  Lett.\/} {\bf 930} L12 (\textit{Preprint} \eprint{2311.08680})

\bibitem{Cardoso:2008bp}
Cardoso V, Miranda A~S, Berti E, Witek H and Zanchin V~T 2009 {\em Phys. Rev.
  D\/} {\bf 79} 064016 (\textit{Preprint} \eprint{0812.1806})

\bibitem{Giri:2022zhf}
Giri S, Nandan H, Joshi L~K and Maharaj S~D 2022 {\em Eur. Phys. J. Plus\/}
  {\bf 137} 181 (\textit{Preprint} \eprint{2204.09006})

\bibitem{Mondal:2020uwp}
Mondal M, Pradhan P, Rahaman F and Karar I 2020 {\em Mod. Phys. Lett. A\/} {\bf
  35} 2050249 (\textit{Preprint} \eprint{2008.11022})

\bibitem{Olmo:2023lil}
Olmo G~J, Rosa J~L, Rubiera-Garcia D and Saez-Chillon~Gomez D 2023 {\em Class.
  Quant. Grav.\/} {\bf 40} 174002 (\textit{Preprint} \eprint{2302.12064})

\bibitem{Ovalle:2016pwp}
Ovalle J, Casadio R and Sotomayor A 2017 {\em Adv. High Energy Phys.\/} {\bf
  2017} 9756914 (\textit{Preprint} \eprint{1612.07926})

\bibitem{Shukla:2024tkw}
Shukla B, Das P~P, Dudal D and Mahapatra S 2024 {\em Phys. Rev. D\/} {\bf 110}
  024068 (\textit{Preprint} \eprint{2404.02095})

\bibitem{Das:2023ess}
Das A, Roy~Chowdhury A and Gangopadhyay S 2024 {\em Class. Quant. Grav.\/} {\bf
  41} 015018 (\textit{Preprint} \eprint{2306.00646})

\bibitem{Du:2024uhd}
Du Y~Z, Li H~F, Ma Y~B and Gu Q 2025 {\em Eur. Phys. J. C\/} {\bf 85} 78
  (\textit{Preprint} \eprint{2403.20083})

\bibitem{Gogoi:2024akv}
Gogoi N~J, Acharjee S and Phukon P 2024 {\em Eur. Phys. J. C\/} {\bf 84} 1144
  (\textit{Preprint} \eprint{2404.03947})

\bibitem{Barreto:2022ohl}
Barreto W and da~Rocha R 2022 {\em Phys. Rev. D\/} {\bf 105} 064049
  (\textit{Preprint} \eprint{2201.08324})

\bibitem{sano1985measurement}
Sano M and Sawada Y 1985 {\em Physical Rev. Lett.\/} {\bf 55} 1082

\bibitem{Cornish:2003ig}
Cornish N~J and Levin J~J 2003 {\em Class. Quant. Grav.\/} {\bf 20} 1649–1660
  (\textit{Preprint} \eprint{gr-qc/0304056})

\bibitem{Bambi:2025wjx}
Bambi C {\em et~al.\/} 2025 {Black hole mimickers: from theory to observation}
  (\textit{Preprint} \eprint{2505.09014})

\bibitem{Casadio:2024lgw}
Casadio R, Kamenshchik A and Ovalle J 2024 {\em Phys. Rev. D\/} {\bf 109}
  024042 (\textit{Preprint} \eprint{2401.03980})

\bibitem{Yang:2023agi}
Yang Y, Liu D, \"Ovg\"un A, Lambiase G and Long Z~W 2024 {\em Phys. Rev. D\/}
  {\bf 109} 024002 (\textit{Preprint} \eprint{2307.09344})

\bibitem{Mazza:2021rgq}
Mazza J, Franzin E and Liberati S 2021 {\em JCAP\/} {\bf 04} 082
  (\textit{Preprint} \eprint{2102.01105})

\bibitem{Ramos:2021jta}
Ramos A, Arias C, Avalos R and Contreras E 2021 {\em Annals Phys.\/} {\bf 431}
  168557 (\textit{Preprint} \eprint{2107.01146})

\bibitem{Daghigh:2022uws}
Daghigh R~G, Green M~D and Morey J~C 2023 {\em Phys. Rev. D\/} {\bf 107} 024023
  (\textit{Preprint} \eprint{2209.09324})

\bibitem{Berti:2009kk}
Berti E, Cardoso V and Starinets A~O 2009 {\em Class. Quant. Grav.\/} {\bf 26}
  163001 (\textit{Preprint} \eprint{0905.2975})

\bibitem{Cardoso:2001bb}
Cardoso V and Lemos J~P~S 2001 {\em Phys. Rev. D\/} {\bf 64} 084017
  (\textit{Preprint} \eprint{gr-qc/0105103})

\bibitem{Konoplya:2011}
Konoplya R~A and Zhidenko A 2011 {\em Rev. Mod. Phys.\/} {\bf 83} 793–836
  (\textit{Preprint} \eprint{1102.4014})

\bibitem{Santos:2019yzk}
Santos E~C, Fabris J~C and de~Freitas~Pacheco J~A 2019  (\textit{Preprint}
  \eprint{1903.04874})

\bibitem{Iyer:1986np}
Iyer S and Will C~M 1987 {\em Phys. Rev. D\/} {\bf 35} 3621

\bibitem{Konoplya:2019hlu}
Konoplya R~A, Zhidenko A and Zinhailo A~F 2019 {\em Class. Quant. Grav.\/} {\bf
  36} 155002 (\textit{Preprint} \eprint{1904.10333})

\bibitem{Churilova:2019jqx}
Churilova M~S 2019 {\em Eur. Phys. J. C\/} {\bf 79} 629 (\textit{Preprint}
  \eprint{1905.04536})

\bibitem{Konoplya:2023moy}
Konoplya R~A and Zhidenko A 2023 {\em Class. Quant. Grav.\/} {\bf 40} 245005
  (\textit{Preprint} \eprint{2309.02560})

\bibitem{carlip2014black}
Carlip S 2014 {\em Int. J. Mod. Phys. D\/} {\bf 23} 1430023 (\textit{Preprint}
  \eprint{1410.1486})

\bibitem{wald2001thermodynamics}
Wald R~M 2001 {\em Living Rev. Rel.\/} {\bf 4} 6 (\textit{Preprint}
  \eprint{gr-qc/9912119})

\bibitem{bekenstein1980black}
Bekenstein J~D 1980 {\em Phys. Today\/} {\bf 33} 24--31

\bibitem{hawking1976black}
Hawking S~W 1976 {\em Phys. Rev. D\/} {\bf 13} 191--197

\bibitem{Feng:2024bsx}
Feng W, da~Rocha R and Casadio R 2024 {\em Eur. Phys. J. C\/} {\bf 84} 586
  (\textit{Preprint} \eprint{2401.14540})

\bibitem{Casadio:2023pmh}
Casadio R, da~Rocha R, Giusti A and Meert P 2024 {\em Phys. Lett. B\/} {\bf
  849} 138466 (\textit{Preprint} \eprint{2310.07505})

\bibitem{Casadio:2022pla}
Casadio R, da~Rocha R, Meert P, Tabarroni L and Barreto W 2023 {\em Class.
  Quant. Grav.\/} {\bf 40} 075014 (\textit{Preprint} \eprint{2206.10398})

\bibitem{Promsiri:2020jga}
Promsiri C, Hirunsirisawat E and Liewrian W 2020 {\em Phys. Rev. D\/} {\bf 102}
  064014 (\textit{Preprint} \eprint{2003.12986})

\bibitem{Dey:2016pei}
Dey A, Roy P and Sarkar T 2018 {\em JHEP\/} {\bf 04} 098 (\textit{Preprint}
  \eprint{1609.02290})

\bibitem{Czinner:2015eyk}
Czinner V~G and Iguchi H 2016 {\em Phys. Lett. B\/} {\bf 752} 306–310
  (\textit{Preprint} \eprint{1511.06963})

\bibitem{Czinner:2017tjq}
Czinner V~G and Iguchi H 2017 {\em Eur. Phys. J. C\/} {\bf 77} 892
  (\textit{Preprint} \eprint{1702.05341})

\bibitem{Abreu:2020vkc}
Abreu E~M~C and Neto J~A 2021 {\em EPL\/} {\bf 135} 10001 (\textit{Preprint}
  \eprint{2009.05012})

\bibitem{nakarachinda2021effective}
Nakarachinda R, Hirunsirisawat E, Tannukij L and Wongjun P 2021 {\em Phys. Rev.
  D\/} {\bf 104} 064003 (\textit{Preprint} \eprint{2106.02838})

\bibitem{ditta2023thermal}
Ditta A, Javed F, Maurya S~K, Mustafa G and Atamurotov F 2023 {\em Phys. Dark
  Univ.\/} {\bf 42} 101345

\bibitem{hawking1975particle}
Hawking S~W 1975 {\em Commun. Math. Phys.\/} {\bf 43} 199--220 [Erratum:
  Commun.Math.Phys. 46, 206 (1976)]

\bibitem{wei2013observing}
Wei S~W and Liu Y~X 2013 {\em JCAP\/} {\bf 11} 063 (\textit{Preprint}
  \eprint{1311.4251})

\bibitem{mashhoon1973scattering}
Mashhoon B 1973 {\em Phys. Rev. D\/} {\bf 7} 2807--2814

\bibitem{ditta2023thermodynamic}
Ditta A, Tiecheng X, Ali R, Atamurotov F, Mahmood A and Mumtaz S 2023 {\em
  Annals Phys.\/} {\bf 453} 169326

\bibitem{ditta2022thermal}
Ditta A, Tiecheng X, Mustafa G, Yasir M and Atamurotov F 2022 {\em Eur. Phys.
  J. C\/} {\bf 82} 756

\bibitem{panah2020charged}
Eslam~Panah B, Jafarzade K and Hendi S~H 2020 {\em Nucl. Phys. B\/} {\bf 961}
  115269 (\textit{Preprint} \eprint{2004.04058})

\bibitem{chen2024quasi}
Chen H, Dong S~H, Hassanabadi S, Heidari N and Hassanabadi H 2024 {\em Chin.
  Phys. C\/} {\bf 48} 085105

\end{thebibliography}

\end{document}